\documentclass[twocolumn]{aastex7} 

\usepackage{pbox}
\usepackage{tabularx}
\usepackage[usestackEOL]{stackengine}
\usepackage{makecell}

\usepackage{mathtools}
\usepackage[T1]{fontenc}
\usepackage{amsmath}
\usepackage{enumitem}

\begin{document}

\def\bluetext#1{{\color{black}#1}}
\def\mgtext#1{{\color{black}#1}}

\title{The Tantalizing Case of the Quasar J0950$+$5128 – I. Presentation of the Data and Detailed Exploration of the Binary Supermassive Black Hole Scenario}

\author[0000-0001-9806-4034]{Niana N. Mohammed}
\affiliation{Department of Astronomy and Astrophysics and Institute for Gravitation and the Cosmos, Penn State University,
525 \\Davey Lab, 251 Pollock Road, University Park, PA 16802, USA}
\affiliation{Fisk University, Department of Life and Physical Sciences, 1000 17th Avenue N, Nashville, TN 37208, USA}
\email{nnm5189@psu.edu}

\author[0000-0001-8557-2822]{Jessie C. Runnoe}
\affiliation{Vanderbilt University, Department of Physics \& Astronomy, 6301 Stevenson Center, Nashville, TN 37235, USA}
\affiliation{Fisk University, Department of Life and Physical Sciences, 1000 17th Avenue N, Nashville, TN 37208, USA}
\email{jessie.c.runnoe@vanderbilt.edu}

\author[0000-0002-3719-940X]{Michael Eracleous}
\affiliation{Department of Astronomy and Astrophysics and Institute for Gravitation and the Cosmos, Penn State University,
525 \\Davey Lab, 251 Pollock Road,  University Park, PA 16802, USA}
\email{mxe17@psu.edu}

\author[0000-0002-7835-7814]{Tamara Bogdanovi\'c}
\affiliation{Center for Relativistic Astrophysics, School of Physics, Georgia Institute of Technology, Atlanta, GA 30332, USA}
\email{tamarab@gatech.edu}

\author[0000-0003-2686-9241]{Daniel Stern}
\affiliation{Jet Propulsion Laboratory, California Institute of Technology,
4800 Oak Grove Drive, Pasadena, CA 91109, USA}
\email{daniel.k.stern@jpl.nasa.gov}

\author[0000-0003-1407-6607]{Joseph Simon}
\altaffiliation{NSF Astronomy and Astrophysics Postdoctoral Fellow}
\affiliation{Department of Astrophysical and Planetary Sciences, University of Colorado, Boulder, CO 80309, USA}
\email{Joseph.Simon-1@colorado.edu}

\author[0000-0003-3579-2522]{Maria Charisi}
\affiliation{Department of Physics and Astronomy, Washington State University, Webster Hall 1245, Pullman, WA 99163, USA}
\affiliation{Institute of Astrophysics, FORTH, GR-71110, Heraklion, Greece}
\email{mariacharisi1@gmail.com}

\author{T. Joseph W. Lazio}
\affiliation{Jet Propulsion Laboratory, California Institute of Technology,
4800 Oak Grove Drive, Pasadena, CA 91109, USA}
\email{joseph.lazio@jpl.nasa.gov}

\author[0009-0000-4083-2547]{Kaitlyn Szekerczes}
\affiliation{Department of Astronomy and Astrophysics and Institute for Gravitation and the Cosmos, Penn State University,
525 \\Davey Lab, 251 Pollock Road, University Park, PA 16802, USA}
\email{kas7882@psu.edu}

\author[0000-0002-8187-1144]{Steinn Sigur{\dh}sson}
\affiliation{Department of Astronomy and Astrophysics and Institute for Gravitation and the Cosmos, Penn State University,
525 \\Davey Lab, 251 Pollock Road, University Park, PA 16802, USA}
\email{sxs540@psu.edu}

\author[0000-0001-7306-1830]{Collin Dabbieri}
\affiliation{Vanderbilt University, Department of Physics \& Astronomy, 6301 Stevenson Center, Nashville, TN 37235, USA}
\email{collin.m.dabbieri@vanderbilt.edu}








\begin{abstract}


Spectroscopic observations of the quasar J0950$+$5128 spanning 22 years {\color{black}reveal monotonic radial velocity variations in its broad H$\beta$ emission line. Moreover, the line profile becomes broader over time,} necessitating careful measurements. We present robust H$\beta$ velocity shift measurements obtained via cross correlation, applied to both the full spectra and to isolated broad H$\beta$ components derived from spectral decomposition. 
{\color{black}We also examine the light curves for variability consistent with the spectroscopic trends. Using Lomb-Scargle periodogram analysis we find no significant periodic signal.}
We consider several interpretations for the observed changes, including a binary supermassive black hole, dust-cloud obscuration, outflows, a recoiling black hole, and a single perturbed, disk-like broad-line region. \mgtext{We deem the binary and perturbed broad-line region scenarios to be physically plausible.} 
\bluetext{The binary interpretation is the only one for which we can immediately compare a physical model to the available data. Thus, we incorporate radial velocity ``jitter'' to emulate typical quasar variability and fit the radial velocity curve with a Keplerian model to examine whether it can reproduce the observations. In this context,} the available observations trace only a segment of the putative orbit. The fit yields a period of 33 years (observed frame) and an eccentricity of 0.65, with lower limits on the semi-major axis and black hole mass of $10^{-2}\;$pc and $10^7\;{\rm M}_\odot$, respectively. {\color{black}Thus, J0950$+$5128} is a \mgtext{binary candidate deserving further study.} 
{\color{black}The single, perturbed broad-line region interpretation remains viable but requires additional observations and modeling for further evaluation.} Continued monitoring is, therefore, essential. 
\end{abstract}



\section{Introduction} \label{sec:intro}

\subsection{Background and Motivation}
\label{subsec:background_motivation}

Galaxy mergers are a frequently observed phenomenon, and given that most massive galaxies have central supermassive black holes (BHs; \citealt{kormendy1995inward}), supermassive black hole binaries (SBHBs) are thought to be an inevitable
consequence of the merger process. 
\cite{begelman1980massive} described how a SBHB evolves in a post-galaxy merger. The evolution begins with the stellar bulges and respective BHs of the two galaxies sinking toward the center of the merger remnant. This process is initially driven by dynamical friction (\citealt{chandrasekhar1943dynamical}), which acts as the dominant mechanism for angular momentum loss for $\sim$10$^9$~yr for typical masses and can bring the BH pair  closer. At a separation of $\lesssim10$~pc, the BHs form 
a gravitationally bound binary (e.g., \citealt{bogdanovic2022electromagnetic}).
Eventually, the binary ``hardens'' as its orbital velocity becomes comparable to the velocity dispersion of the stars. At this point, stellar scattering takes over as the dominant mechanism that further reduces the binary separation. Stellar scattering is effective 
so long as the number of stars with the right orbits to interact with the binary can be replenished (\citealt{milosavljevic2003long, yu2005evolution}). If this reservoir of stars is depleted, the binary separation may slow its evolution or stall at a separation of $\sim1$ pc (known as the ``last parsec problem''). However, several models––ranging from angular momentum loss through interactions with a circumbinary accretion disk or gaseous cloud (e.g., \citealt{armitage2002accretion, escala2004role, dotti_lisa2006, cuadra2009massive}) to dynamical models that consider non-spherical and anisotropic stellar distributions (e.g., \citealt{merritt2004chaotic, berczik2006efficient, khan2012mergers, khan2013supermassive, holley2015galaxy})––now suggest that there are a number of sufficiently efficient mechanisms for continued orbital decay. Once the binary separation reaches $\lesssim\,$10$^{-2}$~pc, 
angular momentum loss via gravitational waves (GWs) becomes an efficient mechanism, leading the two BHs to coalesce within a Hubble time. 

Unambiguous observational evidence for SBHBs has been confined to the early stages of their evolution. Galaxy mergers and widely separated (kilo-parsec and sub-kilo-parsec scale) BH pairs are observed (e.g., \citealt{comerford2009inspiralling,comerford20091,liu2011dual,ellison2017discovery,hou2019active,foord2020second,goulding2019dual,shen2021dual,koss2023dual}), but bound SBHBs have yet to be imaged and resolved directly. An exception is the radio-loud candidate CSO 0402+379, directly imaged through radio interferometry with a projected separation of $\sim$7~pc (\citealt{rodriguez2006compact,bansal2017constraining}). 
The recent evidence of a stochastic GW background by pulsar timing array (PTA) experiments is compatible with the low-frequency signal expected from
a population of SBHBs at the last stages of their evolution (\citealt{agazie2023nanograv,agazie2023binary,antoniadis2023second,reardon2023search,lee2023searching,miles2025meerkat}), 
 strongly suggesting that SBHBs must 
 evolve to merger, emitting GWs in the process.
This is promising for directly resolving nanohertz GWs (corresponding to orbital periods of tens 
of years) from individual SBHBs at sub-parsec separations. The upcoming Laser Interferometer Space Antenna (LISA; 
\citealt{amaro2023astrophysics,colpi2024}) is also expected to detect SBHB mergers, at higher frequencies 
corresponding to somewhat lower BH masses and later stages of SBHB evolution than those detected by the PTAs.

SBHBs have also been invoked to explain a number of galaxy properties, including 
mass deficits in the cores of elliptical galaxies (e.g., \citealt{milosavljevic2002cores,merritt2006cores}), helical or precessing jets (e.g., \citealt{roos1993precessing,romero2000preccessing}), and X-shaped radio sources (e.g., \citealt{merritt2002xshaped,gopalkrishna2003xshaped}).
Finding unambiguous SBHBs at small separations ($\sim$~0.1$-$1~pc) would greatly enhance our understanding of their evolution and their effects on their 
host galaxies. Thus, indirect methods have been developed to search for their observational signatures. 
The two most commonly used methods, which we describe further below, rely on periodic photometric variability or regular spectroscopic (radial velocity) variability of the broad emission lines. In addition to these, a few other methods have been proposed that are appropriate for SBHBs at separations of $\sim 10^{-2}$ -- 1~pc: peculiar profiles and relative intensities of broad emission lines \citep[e.g.,][]{montuori11,montuori12}, proper motion measurements via infrared interferometry \citep[e.g.][]{dexter20}, periodic modulation of the polarization of scattered light \citep[e.g.][]{dotti22}, kinematic signatures in reverberation mapping signals \citep[e.g.,][]{wang18}, or blue-red asymmetric reverberation of the broad emission lines \citep[e.g.][]{dotti23}.

The photometric method involves looking for periodic optical flux variability in the light curves of quasars (e.g., \citealt{graham2015systematic,charisi2016population,liu2016systematic}) under the assumption that either one or both BHs are accreting, and that the accretion rate is regulated on the binary orbital period 
(\citealt{d2013accretion}). Periodic oscillations can also occur due to relativistic Doppler boosting from an orbiting active BH (\citealt{dorazio2023review}).
A prominent photometric SBHB candidate is OJ 287, whose long-term light curve displayed outbursts separated on a 12-year period (e.g., \citealt{valtonen2008Nature,valtonen2012,komossa2023oj27_1,komossa2023oj287_2}).
The use of periodic photometric variations to identify binaries, especially when only a small number of apparent orbital cycles is observed, 
however, is subject to ambiguities (\citealt{liu2018did,witt2022period,robnik2024period}), since quasars exhibit stochastic variability that can appear periodic (\citealt{vaughan2016false,liu2016systematic,barth_stern2018,elbadry2025arXiv}). 

The spectroscopic method looks for the displaced peaks of broad
emission lines relative to the host galaxy rest frame in quasar spectra (e.g., \citealt{gaskell1983quasars,gaskell1996evidence,gaskell1996supermassive})––this is the approach used in this paper. A single-peaked broad emission line that is significantly offset is a possible signature of bulk motion of 
a single active BH and its surrounding gas in a binary. The possibility that the double-peaked emitters of some quasars can arise from systems with two active BHs has been tested and rejected
{\color{black}based on the lack of radial velocity variations that are indicative of orbital motion (\citealt{eracleous1997rejection,liu2016doublepeak,doan2020improved}). Rather, double-peaked emission lines are best explained as originating in a disk-like broad-line region around a single supermassive black hole \citep[see, for example,][for summary of arguments]{eracleous2003radioloud,eracleous2009review}. While the single-peaked offset line method is viable for SBHB searches,} 
it suffers from the ambiguity that offsets of broad emission lines can also be caused by other effects, such as outflows from the accretion disk (e.g., \citealt{popovic2012super}) or large, non-axisymmetric perturbations of the broad-line region (BLR; e.g., \citealt{gezari2007long,lewis2010long,storchi2017double,schimoia2017evolution}). It is therefore necessary to have multiple observations over extended periods of time to evaluate SBHB candidates identified by this method.


In order to select SBHB candidates based on their displaced broad emission lines, 
one approach is to search
for $\gtrsim$1000 km s$^{-1}$ velocity offsets of the broad H$\beta$ emission lines relative to the narrow emission lines in the spectra of quasars (\citealt{gaskell1983quasars,tsalmantza2011systematic,eracleous2012large,decarli2013nature,liu2014constraining}). Using this method, \cite{eracleous2012large} selected 88 
SBHB candidates which have since undergone extensive studies to characterize their nature such as flux variability, radial velocity changes, and implied population properties (\citealt{runnoe2015large,runnoe2017large,pflueger2018likelihood,nguyen2020pulsar}). 
\cite{runnoe2017large} used 
observations through 2011 and found 29 candidates that displayed steady, and in some cases, statistically significant radial velocity variations. Among these, three objects displayed monotonic changes consistent with the expected behavior of SBHBs. These objects became prime targets for further follow-up observations.
After obtaining new observations over several years using ground-based telescopes, we identify J095036.75$+$512838.1 (hereafter J0950) as a promising SBHB candidate. 

\subsection{Possible Interpretations of a Shifted Broad Emission Line and Scope of this Paper}\label{subsec:possible_interp}

{\color{black}
Here we explore several interpretations for the apparent broad emission line shift of J0950. These include: (a)~a binary supermassive black hole interpretation, which is the hypothesis originally considered in previous works on this object (Section~\ref{subsec:background_motivation}),
(b)~a dust cloud outside the BLR obscuring parts of it as it moves across the line of sight, thereby producing asymmetric broad emission lines that mimic radial velocity shifts; (c)~outflows, where the delays in the arrival times of ejecta moving towards and away from the observer result in apparent blueshifts and redshifts of the broad emission lines across different epochs; (d)~a recoiling BH that has been kicked out of, and subsequently falls back into, the gravitational potential well of its host galaxy; and (e)~a perturbation in a disk-like BLR that can produce a broad emission line peak that shifts in velocity as the perturbation precesses. We revisit these interpretations for further consideration in the Section~\ref{subsec:assess_interp}. 
}

{\color{black} 
The goal of this paper is to present the available data and a thorough study of the variability of the broad H$\beta$ line, 
and to explore the SBHB interpretation in detail. 
It is the first in a series of papers that will investigate possible scenarios for the behavior of this object through observational tests and/or modeling. 
In Section~\ref{sec:properties}, we present the properties of the target and available spectroscopic and photometric data.
Section~\ref{sec:methods} describes the methods for the velocity measurements, simulations of the observed profile change of the broad H$\beta$ emission line, and the relative velocities resulting from the H$\beta$ shift measurements. In Section~\ref{sec:lightcurves}, we present the analysis of the light curves.
We assess the different interpretations of the observed behavior in J0950's spectra and provide a detailed discussion of the SBHB interpretation in Section~\ref{sec:conclusion}.
{\color{black}Section~\ref{sec:conclusions2} summarizes our results and notes future work.}}
We assume a standard cosmology with $H_0=69.6\,\mathrm{km\,s^{-1} Mpc^{-1}}$, $\Omega_M=0.286$, and $\Omega_{\Lambda}=0.714$ \citep{bennett_cosmology2014} throughout this work.\footnote{We calculated luminosity distance using Ned Wright's cosmology calculator \citep{wright2006cosmology}, available at \url{https://www.astro.ucla.edu/~wright/CosmoCalc.html}}

\begin{deluxetable*}{lccccccc}
\tablewidth{0pt} 
\tablecaption{Spectroscopic Observations of J0950
\label{tab:observations}}
\tablehead{
\colhead{Telescope $+$} & \colhead{Aperture} & \colhead{ Observation} & \colhead{S/N$^c$}\\
\colhead{Instrument$^a$} & ~ & \colhead{Date (UT)} & \colhead{}
} 
\startdata 
Sloan$+$SDSS & d=3$^{\prime\prime}$ fiber & 2002.05.15 & 36
\\
HET$+$LRS & $1\overset{\prime\prime}{.}5$ slit& 2010.05.08 & 42 
\\
HET$+$LRS & $1\overset{\prime\prime}{.}5$ slit& 2011.04.08 & 33 
\\
Sloan$+$BOSS & d=2$^{\prime\prime}$ fiber & 2015.01.20 & 33 
\\
Pal$+${DBSP} & $1\overset{\prime\prime}{.}5$ slit& 2019.04.09 & 30 
\\
HET$+$LRS2 & $5\overset{\prime\prime}{.}2^b$ & 2020.02.02 & 19 
\\
HET$+$LRS2 & $4\overset{\prime\prime}{.}2^b$ & 2020.04.02 & 19 
\\
HET$+$LRS2 & $4\overset{\prime\prime}{.}3^b$ & 2021.04.05 & 36 
\\
Keck$+$LRIS & $1\overset{\prime\prime}{.}5$ slit& 2021.04.13 & 51 
\\
HET$+$LRS2 & $4\overset{\prime\prime}{.}6^b$ & 2022.03.25 & 36 
\\
HET$+$LRS2 & $5\overset{\prime\prime}{.}1^b$ & 2022.04.03 & 21 
\\
HET$+$LRS2 & $6\overset{\prime\prime}{.}3^b$ & 2022.12.31 & 28 
\\
HET$+$LRS2 & $4\overset{\prime\prime}{.}9^b$ & 2024.03.30 & 28 
\\
\enddata
\tablecomments{$^a$ 
     SDSS = Sloan Digital Sky Survey spectrograph,
     LRS = Low Resolution Spectrograph, BOSS = Baryon Oscillation Spectroscopic Survey spectrograph,
     DBSP = Double Spectrograph. \\
     $^b$ The effective aperture size used in spectral extraction. LRS2 adjusts the fiber contributions based on an aperture set to 2.5 times the seeing value. \\
     $^c$ The signal-to-noise ratio (S/N), estimated in the continuum around 
     5460~Å.}
\end{deluxetable*}

\begin{figure*}
  \centering  
  \includegraphics[scale=0.50, angle=0]{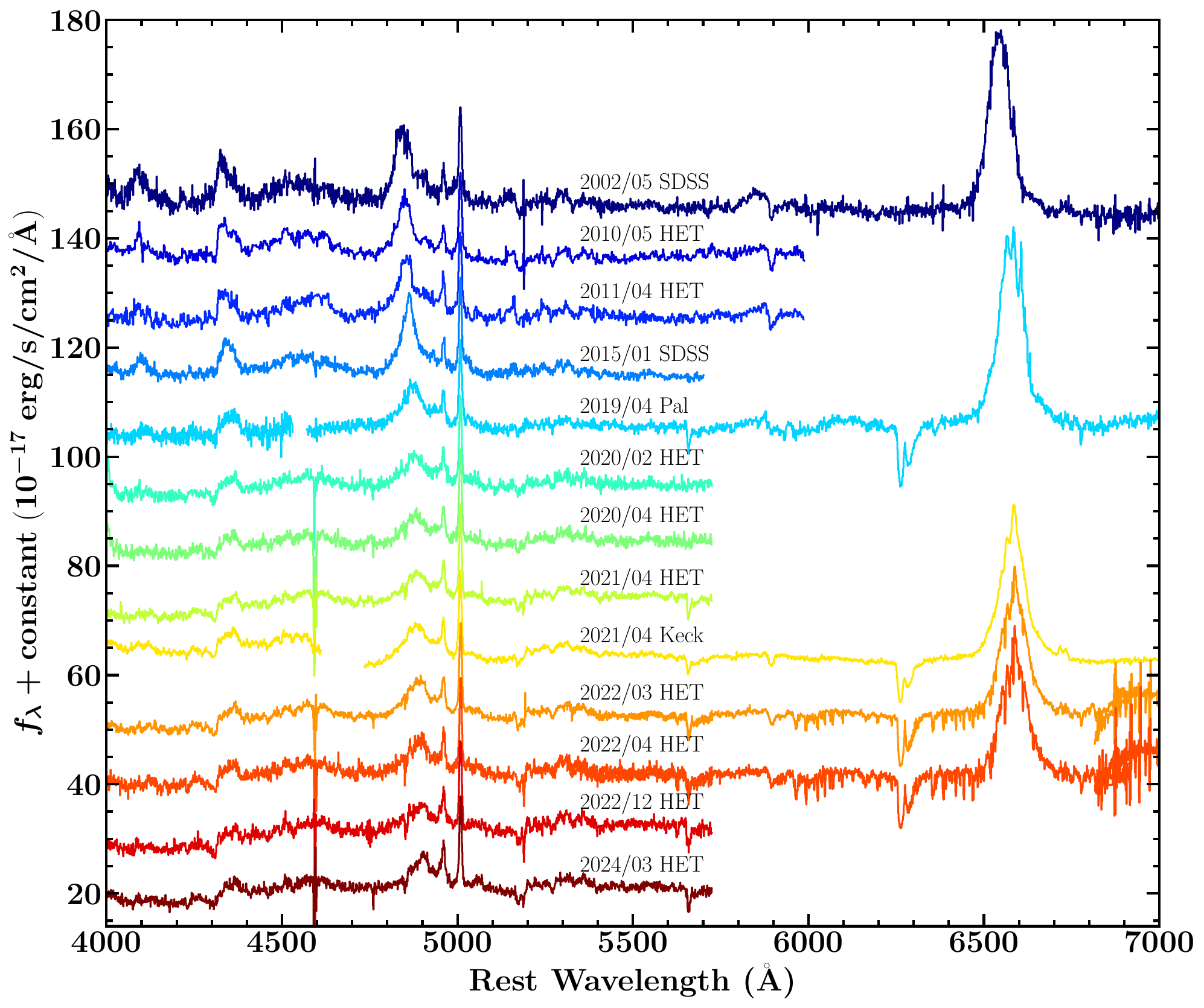}
  \caption{\color{black} The full observed spectra of J0950 spanning approximately 22~years, shown after all the standard calibrations and corrections
  described in Section~\ref{subsec:available_spectra} have been applied. For visual clarity, each spectrum was linearly scaled to match the [O~\textsc{iii}] flux and continuum of the 2002 spectrum, then vertically offset by a constant amount from the last spectrum. The sharp feature seen at a rest wavelength of 4600~Å is the residual from imperfect correction of the [O~\textsc{i}]~$\lambda$5577 telluric emission line. The absorption lines near rest wavelengths of 5650 and 6260~Å are also telluric (due to molecular Oxygen transitions).}
  \label{fig:complete_spectra}
\end{figure*}

\begin{figure*}
  \centering
  \includegraphics[scale=0.45, angle=0]{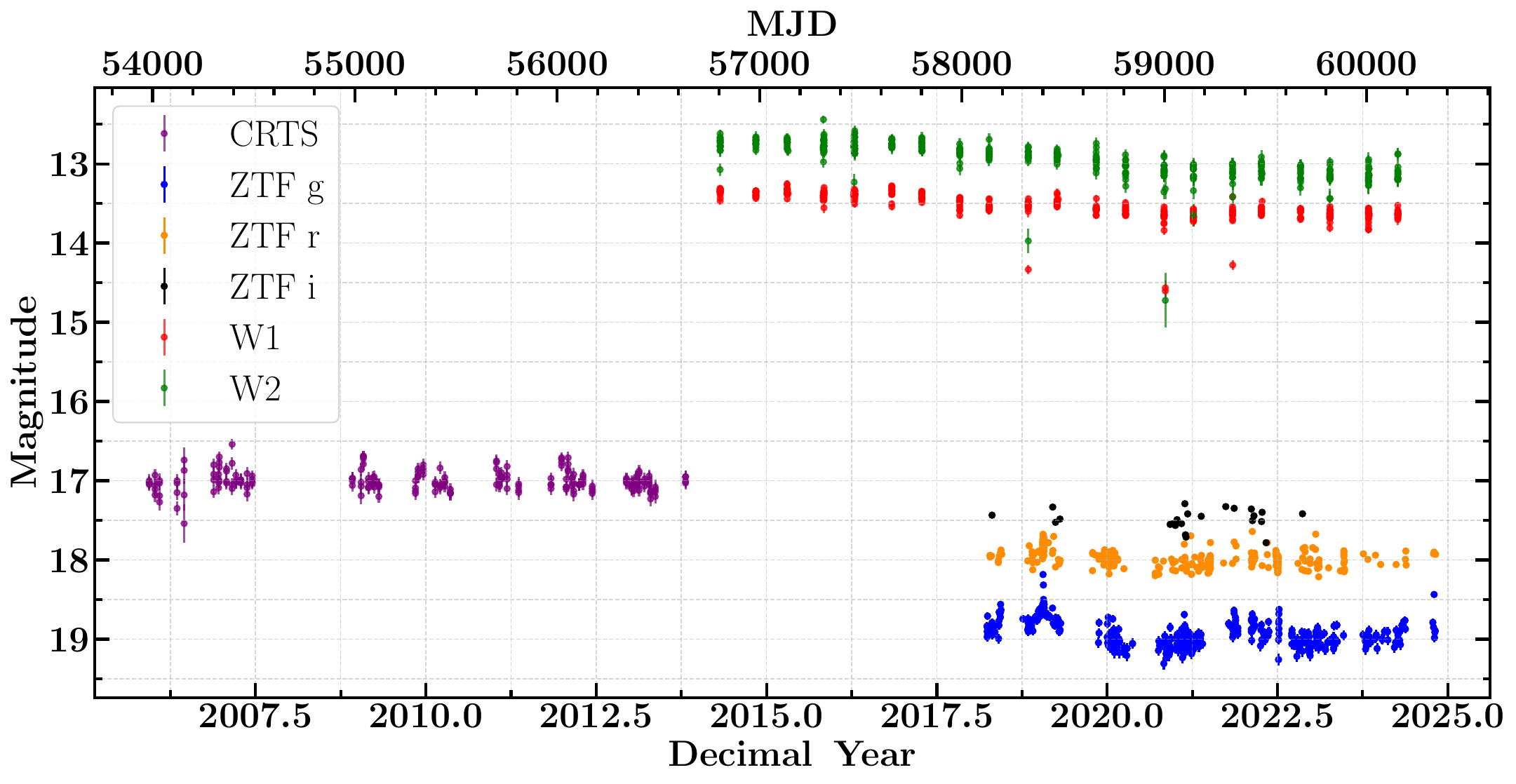}
  \caption{\color{black}The light curves of J0950 from the Zwicky Transient Facility's (ZTF) \textit{g}-, \textit{r}-, and \textit{i}-band, the Catalina Real-Time Transient Survey (CRTS), and the Near-Earth Object Wide-Field Infrared Survey Explorer's (NEOWISE) W1 and W2 filters. 
}
  \label{fig:light_curve}
\end{figure*}

\section{Properties of the Target and Available Data}\label{sec:properties}

\subsection{Properties of the Target}\label{subsec:properties}

J0950 is one of 88 $z\lesssim 0.7$ SBHB candidates selected by \cite{eracleous2012large} from a sample of 15,900 Sloan Digital Sky Survey (SDSS) quasars. The candidates were identified via velocity offsets of their broad H$\beta$ emission lines from the redshift defined by their [O \textsc{iii}] $\lambda\lambda$4959,5007 narrow lines.
Its redshift, measured from the rest wavelength of [O~\textsc{iii}] $\lambda$5007, is $z=0.2144$, which implies a luminosity distance of 1067.9 Mpc. Its absolute and apparent \textit{V}-band magnitudes are $-$21.75 and 18.35 respectively.
The short-term flux variability properties of the continuum and broad H$\beta$ emission of J0950 are much like those observed in normal quasars in the same redshift and luminosity range (\citealt{runnoe2015large}), making it appear to be a normal quasar apart from its offset broad-line. 


\subsection{Available Spectra}\label{subsec:available_spectra}

The current spectroscopic dataset for J0950 consists of spectra taken with the SDSS 2.5m telescope, the Palomar Hale 5m Telescope (Pal), the Keck 10m telescope, and the Hobby-Eberly 11m Telescope (HET) on the dates shown in Table~\ref{tab:observations} and Figure~\ref{fig:complete_spectra}. The calibration process for the early spectra is described in Section 5.1 of \cite{eracleous2012large} and summarized in Section 2.1 of \cite{runnoe2015large}. The recent spectra from the HET were taken with the second-generation Low-Resolution Spectrograph (LRS2; \citealt{lrs2_2016SPIE.9908E..4CC}), which is fed by a fiber bundle. The LRS2 spectra were reduced and calibrated with the \texttt{Panacea}\footnote{\url{https://github.com/grzeimann/Panacea}} package, written by Greg Zeimann. The reduction consists of bias and dark subtraction, fiber tracing and extraction as well as wavelength calibration, fiber-to-fiber normalization, source detection and extraction, and flux calibration.
Starting with those spectra, we carried out the following corrections:

\begin{itemize}
    \item The observed flux density of each spectrum was divided by the 
    continuous atmospheric extinction curve to recover the spectrum above the Earth's atmosphere.
    
    \item The flux density was corrected for Galactic interstellar extinction using the \cite{schlegel1998maps} dust maps\footnote{The dust map was accessed using \texttt{sfdmap} (codebase: \url{https://github.com/kbarbary/sfdmap}) and \texttt{extinction} (\citealt{barbary_k_2016}) \textit{python} packages.} recalibrated by \cite{schlafly2011measuring}, and the extinction law of \cite{fitzpatrick1999correcting}.

    \item The wavelength scale of each spectrum was converted from air to vacuum using the formula provided by \cite{key}, which is an inverse transformation of the vacuum-to-air conversion given by \cite{morton2000atomic}.
    
    \item To verify the absolute wavelength calibration (hence the velocity stability) of the spectra, we used the [O~\textsc{iii}] $\lambda\lambda$4959,5007 doublet, which originates on large scales and serves as an internal wavelength standard. We found that, relative to the 2002 SDSS spectrum the subsequent spectra had shifts between 
    $-72$ and $+41$ km s$^{-1}$,
    as determined using the cross-correlation method described in Section~\ref{sec:methods}. We corrected these shifts to adjust the spectral alignment. The residual uncertainty after the alignment process corresponds to $\pm30$ km s$^{-1}$, which is set by how well we can measure the shift in the [O~\textsc{iii}] doublet via cross-correlation.  
\end{itemize}

\subsection{Available Light Curves}
{\color{black}
J0950 has archival optical and infrared (IR) light curves from 
the Catalina Real-Time Transient Survey (CRTS; \citealt{drake2009crts}), the Near-Earth Object Wide-Field Infrared Survey Explorer (NEOWISE; \citealt{wright2010wise, mainzer2014wise}), and the Zwicky Transient Facility (ZTF; \citealt{masci2019ztf,bellm2019ztf,graham2019ztf}). Figure~\ref{fig:light_curve} shows the light curves from these facilities and in their respective filters. 
We obtained the CRTS light curves (unfiltered visible light observations) from the second data release of The Catalina Surveys.\footnote{
CRTS data access:~\url{http://nesssi.cacr.caltech.edu/DataRelease/CSDR2.html}
} Both NEOWISE (W1 and W2 filters) light curves and ZTF (\textit{g}-, \textit{r}-, and \textit{i}-band) light curves from the 23rd data release were retrieved from the NASA/IPAC Infrared Science Archive (IRSA) archive.\footnote{
NEOWISE data access: \url{https://doi.org/10.26131/IRSA144} \\
ZTF data access: \url{https://doi.org/10.26131/IRSA598}
} 
The light curves span approximately 19 years in total. However, the segments covered by any single instrument are typically shorter than 10 years. 
}


\section{Analysis of Spectra}\label{sec:methods}
{\color{black} We aimed to measure velocity shifts of the broad H$\beta$ line in J0950 across different epochs. To do this, we used two main approaches: (a) cross correlating the H$\beta$ profiles directly in the observed spectra and (b) fitting the spectra with parametric models and cross correlating those in the models. In both cases, we also isolated the broad H$\beta$ line after decomposing the spectra to measure its velocity shifts independently of other spectral features. This required performing a spectral decomposition to remove contaminating components as needed. Our procedure yielded four distinct sets of relative velocity shifts of the broad H$\beta$ lines. 

In Section~\ref{subsec:spectral_decomp}, we describe the spectral decomposition and provide central moments and line properties measured from the model broad H$\beta$ lines. Section \ref{subsec:crosscor} outlines the cross-correlation technique and the resulting velocity shift measurements in both the data and models. In Sections \ref{subsec:non_breathing} and \ref{subsec:breathing}, we perform simulations to investigate the effects of the observed variability of the broad H$\beta$ profile on the shift measurements. Section \ref{subsec:rv_curve} summarizes the different sets of relative velocity shifts and explains which measurements we consider reliable and adopt for the subsequent radial velocity analysis. Throughout the above analyses we describe how we carry out experiments
to determine uncertainties and, in Section~\ref{subsec:uncertainties} we summarize those uncertainties.}

\subsection{Spectral Decomposition and Measurements of the Broad H$\beta$ Line Profiles}\label{subsec:spectral_decomp}

\begin{figure*}
  \centering
  \includegraphics[scale=0.55, angle=0]{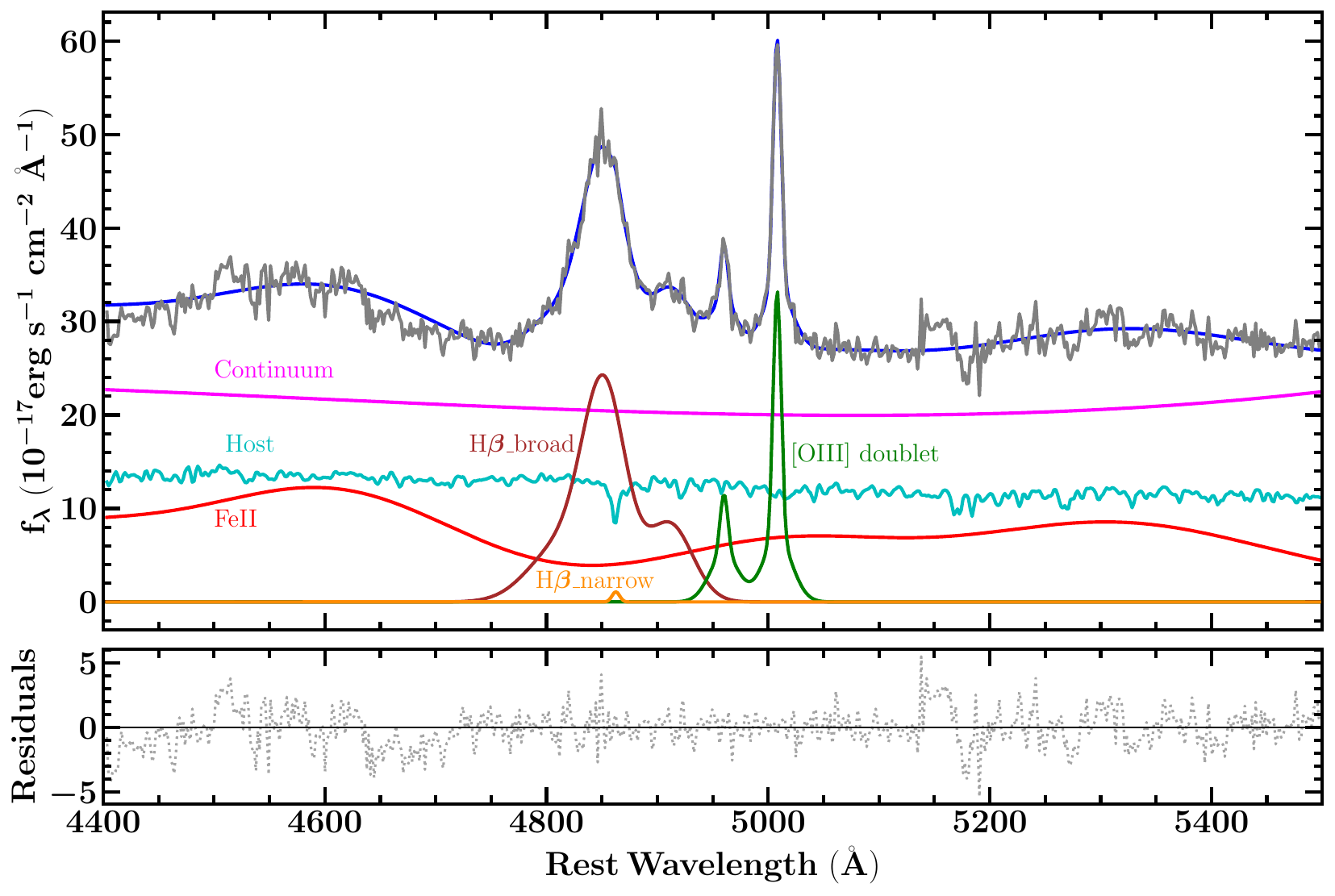}
  \caption{Example spectral decomposition of J0950's 2010 spectrum performed using \texttt{PyQSOfit} \citep{guo2018pyqsofit}. {\color{black}The observed spectrum after subtracting the host galaxy component (cyan), is shown in grey, with the best-fitting model overplotted in blue. The residuals are presented as a light-grey dotted line in the lower panel.} 
  The separate components of the model, including the continuum, 
  Fe~\textsc{ii} lines, broad H$\beta$, and narrow H$\beta$ and [O \textsc{iii}] lines, are also shown.}
  \label{fig:decomp example}
\end{figure*}

We used the \texttt{PyQSOfit} quasar spectrum fitting code (\citealt{guo2018pyqsofit}) to model the spectra and decompose the host galaxy starlight spectrum, quasar continuum, Fe~\textsc{ii} lines, broad and narrow He~\textsc{ii} $\lambda$4687 and H$\beta$ $\lambda$4863 emission lines, and the [O \textsc{iii}] $\lambda\lambda$4959,5007 doublet. {\color{black}The He~\textsc{ii} feature is very weak in all the spectra and, although included in the fits, contributes negligibly to the overall models. To perform the decomposition, we restricted the spectral range to 4400–5500~Å and 
used only the least-squares routine for fitting. Each narrow emission line was modeled with a single Gaussian component, with the centroids and widths of the narrow lines tied together, reflecting their common narrow-line region origin. The broad H$\beta$ line was modeled with four Gaussian components to capture any profile asymmetries and to better fit the less well-defined profiles. The [O \textsc{iii}] wing components were modeled with a Gaussian each and tied to each other in both velocity and width. {\color{black}The width of the Fe~\textsc{ii} emission was allowed to vary between 1200–18,000~km~s$^{-1}$.} 
An example decomposition of the 2010 HET spectrum is shown in Figure~\ref{fig:decomp example}. }

{\color{black}
The resulting models are shown in Appendix~\ref{appendixA}, Figure~\ref{fig:full_model}. 
From these models, we isolated the model broad H$\beta$ profiles, as shown in Figure~\ref{fig:hbeta_model} (left panel). 
Starting from the $n\mathrm{th}$ central moment of a line profile, $\mu_{\mathrm n}=K\sum_i(\lambda_i - \langle\lambda\rangle)^n f_i$, where $\lambda_i$ and $f_i$ are the individual wavelength and flux densities, and $K=1/\sum f_i$ is a normalization constant, we calculated several line-shape metrics. The centroid was computed as $\langle\lambda\rangle=K\sum_i \lambda_i f_i$. From this, we determined the skewness coefficient, $s=\mu_3/\mu_2^{3/2}$, and the kurtosis coefficient, $k=\mu_4/\mu_2^2$. We report these quantities along with the FWHM and integrated flux in Table~\ref{tab:moments}.
To quantify the uncertainties, we selected the model of the J0950 spectrum with median S/N (2015 HET spectrum; see Table~\ref{tab:observations}), and generated 30 realizations by adding Gaussian noise with a constant standard deviation set by the measured S/N. For each realization, we modeled the broad H$\beta$ line 
and measured its central moments and line properties. We report the standard deviation of these measurements across the realizations as representative uncertainties for the central moments and line properties of the broad H$\beta$ line in all spectra of J0950.
}

{\color{black}The broad H$\beta$ profile recovered from a decomposition can be sensitive to the Fe~\textsc{ii} fit. To assess this effect, we varied the width of the Fe~\textsc{ii} lines in the decomposition process and examined its impact on the modeled broad H$\beta$ profile. We selected a subset of spectra that represent the range of observed broad H$\beta$ profiles of J0950. For each spectrum, we fixed the Fe~\textsc{ii} width to be narrower or broader by up to $\pm$~5000~km~s$^{-1}$ from the best-fit value, ensuring that each fixed width provided a reasonable fit while allowing all other parameters to vary. In each case, we omitted the model Fe~\textsc{ii} and all other spectral features from the full model spectrum to isolate the broad H$\beta$ profile. We describe the implications of this Fe~\textsc{ii} modeling experiment in Section~\ref{subsec:crosscor_models}, where we present the methodology for shift measurements.}

\begin{deluxetable*}{lccccccc} 
\tablewidth{0pt} 
\tablecaption{Central moments and properties of the broad H$\beta$ line profiles
\label{tab:moments}}
\renewcommand{\arraystretch}{1.2} 
\tablehead{
\colhead{Observation} & \colhead{Centroid$^{+200}_{-200}$} & 
\colhead{Skewness$^{+0.1}_{-0.1}$} & 
\colhead{Kurtosis$^{+0.2}_{-0.2}$} & 
\colhead{FWHM$^{+60}_{-60}$} &  
\colhead{Integrated$^{+0.3}_{-0.3}$}\\ 
\colhead{Date (UT)} & \colhead{(km s$^{-1}$)$^a$} & \colhead{Coefficient} & 
\colhead{Coefficient}
& \colhead{(km s$^{-1}$)} & \colhead{Flux $^b$}
}
\startdata
2002.05.15 & $-670$ & $-1.14$ & $1.37$ & $3100$ & 7.50 \\
2010.05.08 & $260$ & $-1.19$ & $1.66$ & $3540$ & 10.10 \\
2011.04.08 & $160$ & $-1.26$ & $1.75$ & $3800$ & 9.70 \\
2015.01.20 & $540$ & $-1.09$ & $2.24$ & $3310$ & 8.80 \\
2019.04.09 & $1690$ & $-1.22$ & $1.67$ & $4560$ & 11.80 \\
2020.02.02 & $1020$ & $-1.06$ & $2.65$ & $5160$ & 9.70 \\
2020.04.02 & $980$ & $-1.17$ & $2.50$ & $5240$ & 9.00 \\
2021.04.05 & $1460$ & $-1.17$ & $2.67$ & $5990$ & 9.50 \\
2022.03.25 & $860$ & $-1.41$ & $2.92$ & $5530$ & 9.10 \\
2022.04.03 & $1750$ & $-0.98$ & $2.78$ & $7240$ & 11.80 \\
2022.12.31 & $2020$ & $-1.08$ & $2.68$ & $8330$ & 9.10 \\
2024.03.30 & $1830$ & $-1.06$ & $2.87$ & $5760$ & 9.00
\\
\enddata
\tablecomments{
    $^a$ Centroid shifts are expressed relative to the rest wavelength of H$\beta$.
    \\
    $^b$ Integrated fluxes are in units of $10^{-15}$ erg s$^{-1}$ cm$^{-2}$.
    }    
\end{deluxetable*}

\begin{figure}
  \includegraphics[width=\columnwidth, angle=0]{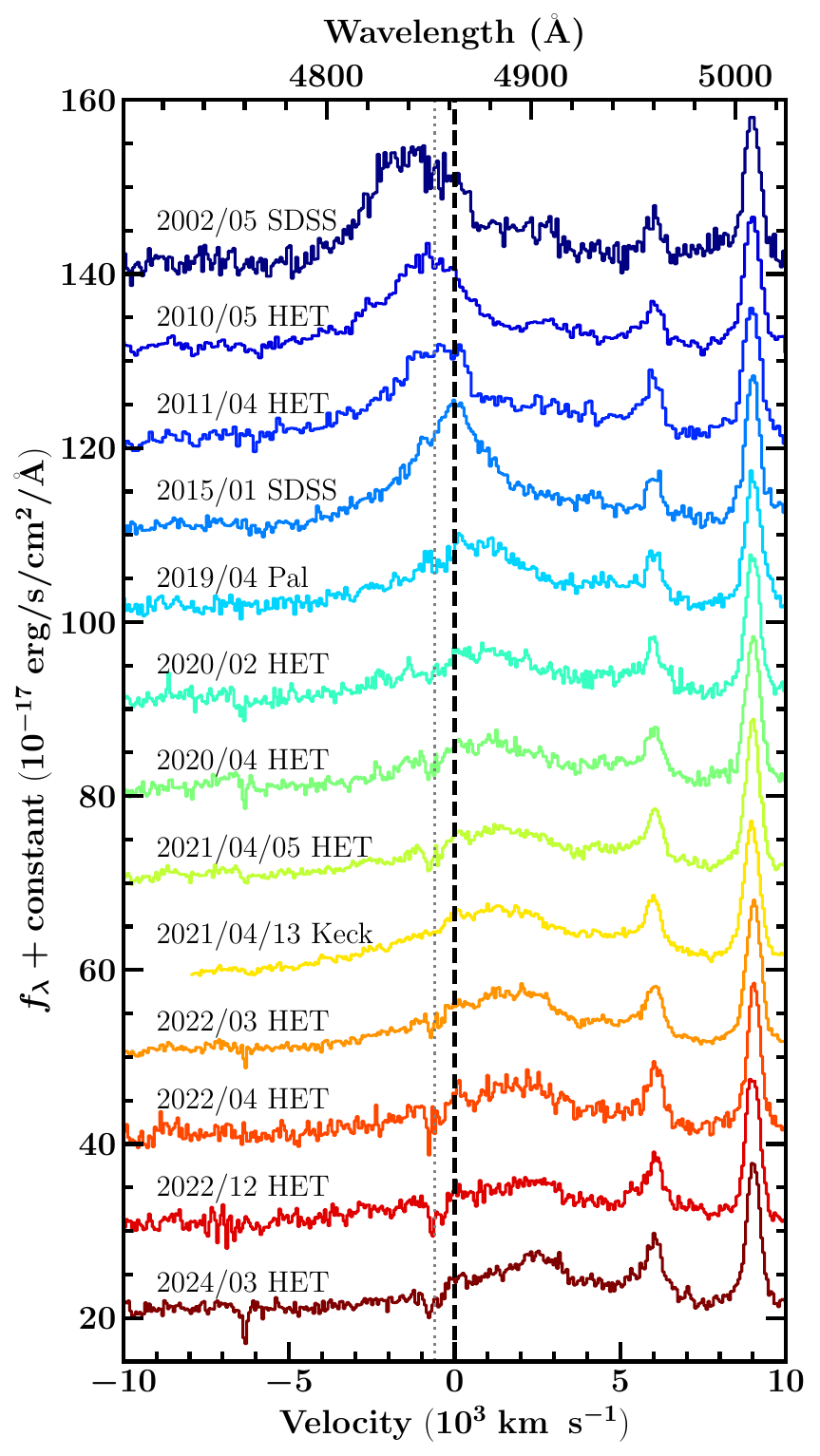}
  \caption{The spectra of J0950 obtained over 22 years, {\color{black}identical to those in Figure \ref{fig:complete_spectra} but shown over a narrower wavelength range and on a velocity scale}. 
The broad H$\beta$ line is Doppler shifted to bluer wavelengths at first, and becomes redshifted in the later spectra. The line shape also displays variability over time. The rest wavelength of H$\beta$ (vertical black dashed line) is set by the redshift measured from the [O~\textsc{iii}] $\lambda\lambda$4959,5007 lines emitted from larger scales in the host galaxy. The vertical grey dotted line at 4853~Å ($-597\,\mathrm{km\,s^{-1}}$) marks the location of the wavelength of the Na I D emission line from the night sky. 
This line is not always corrected properly; therefore, the residual from Na~I~D subtraction in some spectra has been masked out in our later analysis.}
  \label{fig:data}
\end{figure}

\begin{figure}[htb!]
  \includegraphics[width=\columnwidth, angle=0]{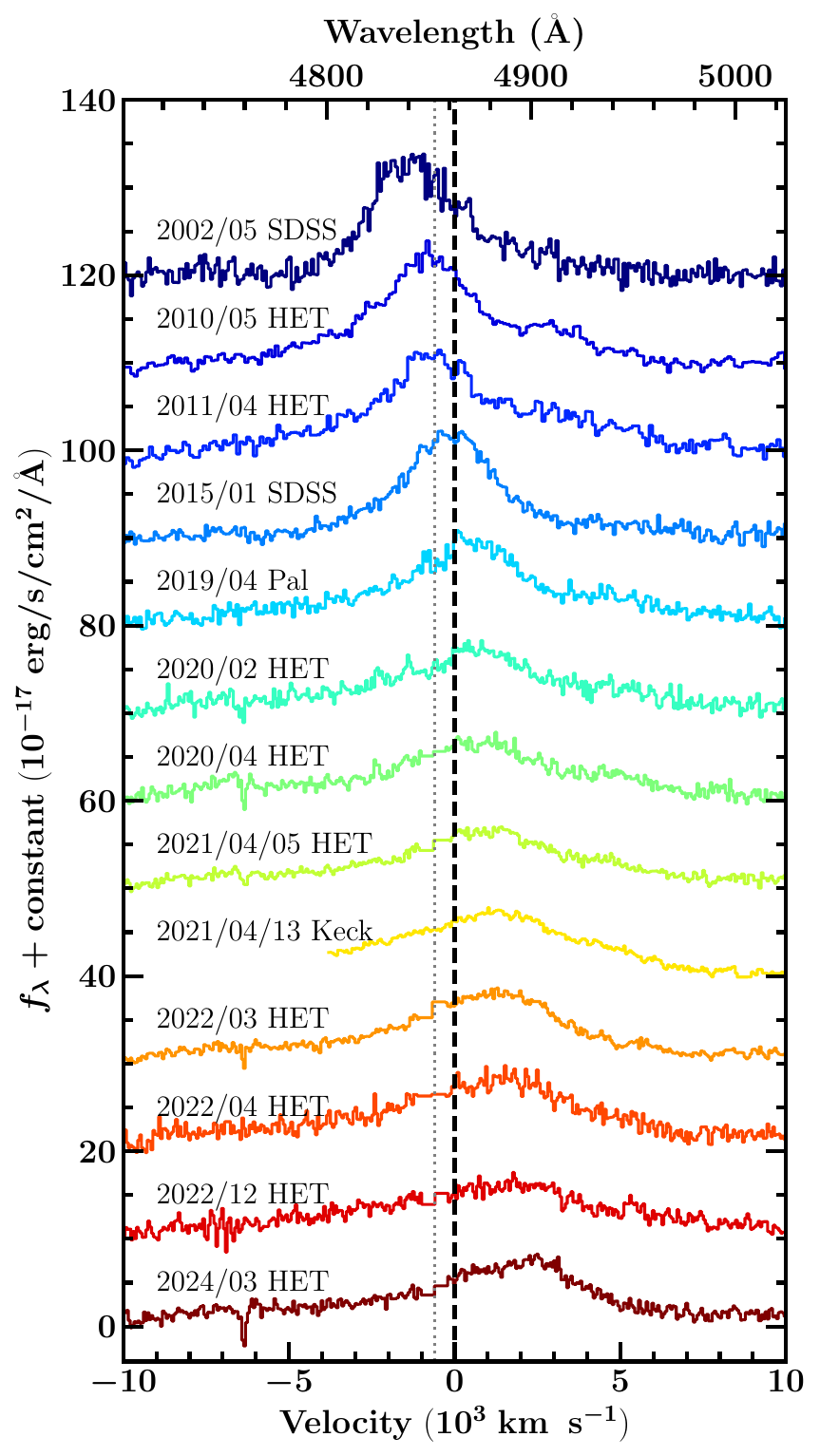}
  \caption{The broad H$\beta$ emission of J0950 isolated by spectral decomposition (the continuum, host, Fe~\textsc{ii} lines, broad \& narrow He~\textsc{ii}, narrow H$\beta$, and [O \textsc{iii}] doublet have been subtracted out, {\color{black}as described in Section \ref{subsec:crosscor_data}). The lines are vertically offset by a constant amount.} See Figure~\ref{fig:data} caption for further details.
  }
  \label{fig:decomp spectra}
\end{figure}

\subsection{Measuring Broad H$\beta$ Velocity Shifts} \label{subsec:crosscor}
The technique used throughout this work to measure shifts of the broad H$\beta$ emission lines of J0950 is the cross-correlation method described in \cite{eracleous2012large} \citep[see also][]{shen2013constraining}. In summary, the method involves first taking two spectra, and linearly scaling one spectrum to match the other in its broad-line flux and continuum. The linear scaling minimizes the difference between the spectra so that the velocity offsets of their broad lines may be compared faithfully. To measure the relative broad H$\beta$ shift, one spectrum is held fixed while the other is progressively shifted in wavelength, and the $\chi^2$ at each shift is calculated in a window containing only the H$\beta$ profile. The shift value corresponding to the minimum $\chi^2$ is then adopted. Figure 9 in \cite{eracleous2012large} demonstrates applications of the cross-correlation method. 

In this work, each broad H$\beta$ velocity shift 
represents the average of two relative shift measurements between broad H$\beta$ profiles: one with the 2002 spectrum profile held constant while the profile of a later spectrum was shifted for cross-correlation, and another with the later spectrum held constant. Below we describe how we cross correlated the broad H$\beta$ lines in both the data and corresponding models 
to quantify sources of random and systematic errors.

\begin{figure*}
  \centering
  \includegraphics[scale=0.55, angle=0]{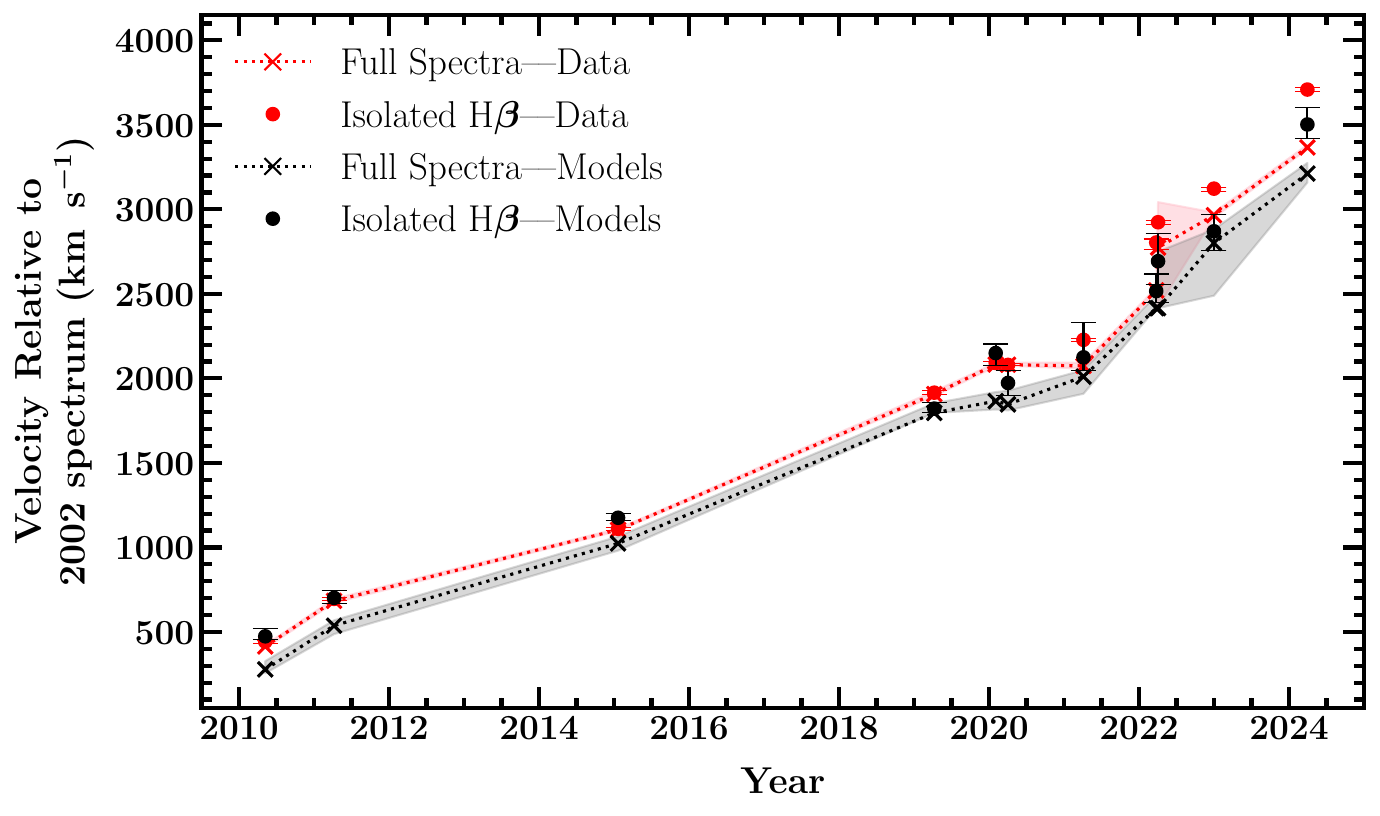}
  \caption{
Relative velocity shifts obtained by cross-correlating the broad H$\beta$ emission lines of J0950 in both the data and their models. The shifts are expressed relative to the 2002 spectrum (itself not plotted). Four sets of measurements are shown
with different markers. 
{\color{black}Velocities from the full observed spectra (Figure~\ref{fig:data}) are shown as red $\times$s connected by a dotted line, and those from the isolated broad H$\beta$ (Figure~\ref{fig:decomp spectra}) as red circles.}
{\color{black} Black $\times$s, connected by a dotted line, and black circles correspond to measurements from models of the full spectra and the isolated H$\beta$, respectively (see Appendix~\ref{appendixA}).}
Both shaded regions and error bars represent 68\% confidence intervals {\color{black}and correspond to formal cross-correlation uncertainties only}. {\color{black}
Additional measurement uncertainties of the same magnitude, not shown in this plot, come from spectral decomposition (especially the subtraction of the Fe~\textsc{ii} complex) and the increasing width of the H$\beta$ line with time {\color{black}(see Section~\ref{subsec:uncertainties})}.} 
}
  \label{fig:rad_vel_both}
\end{figure*}

\subsubsection{ Broad H$\beta$ Velocity Shifts in the Data} \label{subsec:crosscor_data}

We measured velocity shifts of the broad H$\beta$ using two different approaches and compared the results. First, we cross correlated the full spectra in Figure~\ref{fig:data}, following the same procedure in \cite{eracleous2012large}{\color{black}, after modeling and subtracting the host galaxy contribution with \texttt{PyQSOfit} (\citealt{guo2018pyqsofit})}. In the second approach, we cross correlated the \textit{isolated} broad H$\beta$ emission lines (Figure~\ref{fig:decomp spectra}), obtained by following the spectral decomposition method adopted by other works (\citealt{shen2013constraining,shen2019sloan,guo2019constraining}). {\color{black}To isolate the broad H$\beta$ profiles, we first fit the spectra of J0950 as described in Section~\ref{subsec:spectral_decomp}. After fitting, all components except the broad H$\beta$ were subtracted from each spectrum in Figure~\ref{fig:data} to get the isolated broad H$\beta$ profiles shown in Figure~\ref{fig:decomp spectra}.}

The goal of measuring shifts using the full spectra and the isolated broad H$\beta$ emission lines is to check if the two methods yield consistent results. 
{\color{black}Figure~\ref{fig:rad_vel_both} shows the relative velocities measured from the full spectra (red dotted line) and the isolated H$\beta$ lines (red circles), along with shaded regions and error bars representing 68\% confidence intervals.}
{\color{black}We excluded the shift measurement from the 2021 Keck spectrum because a piece of the continuum near the broad H$\beta$, which is essential for reliable spectral decomposition and cross-correlation, is missing due to the dichroic cutoff. Nonetheless, excluding this spectrum has a negligible effect as we have a measurement from an HET spectrum taken at approximately the same time.}

As a consistency check, we applied the measured H$\beta$ shifts to four of our spectra that also include broad H$\alpha$ {\color{black}(see Figure \ref{fig:complete_spectra})} and verified {\color{black}by visual inspection} that those shifts bring the broad H$\alpha$ profile into proper alignment. {\color{black} We do not use H$\alpha$ as a kinematic tracer because it cannot be observed with H$\beta$ in a single exposure with our primary telescope, the HET, and because H$\alpha$ at the redshift of J0950 is contaminated by strong sky lines that ultimately reduce the S/N and undermine our ability to use the narrow lines in its vicinity as a velocity standard.} As a second consistency check, we cross-correlated successive spectra and verified that we obtained a shift equivalent to the difference of shifts to the first spectrum.
{\color{black} Interestingly, among the observed spectra obtained within two months or less of each other in 2020 and 2022 (see Figure \ref{fig:data}), there is a velocity difference of $\sim$80~km~s$^{-1}$ between the two spectra in 2022. Such random velocity variations on short timescales are often caused by subtle changes in the broad H$\beta$ line profile, which can be difficult to discern by eye. These velocity fluctuations are manifestations of radial velocity ``jitter'', which we discuss further in Sections~\ref{subsec:uncertainties} and \ref{subsec:rv_fitting}.}

While the relative velocity shifts from the two methods increase uniformly, 
{\color{black}the shifts from cross-correlating the isolated H$\beta$ are systematically higher than those from the full spectra at higher velocities.}
This difference is primarily an artifact caused by the presence of Fe~\textsc{ii} emission lines in the full spectra, as discussed in Section~\ref{subsec:crosscor_models}. 

\begin{figure}
  \includegraphics[width=\columnwidth, angle=0]{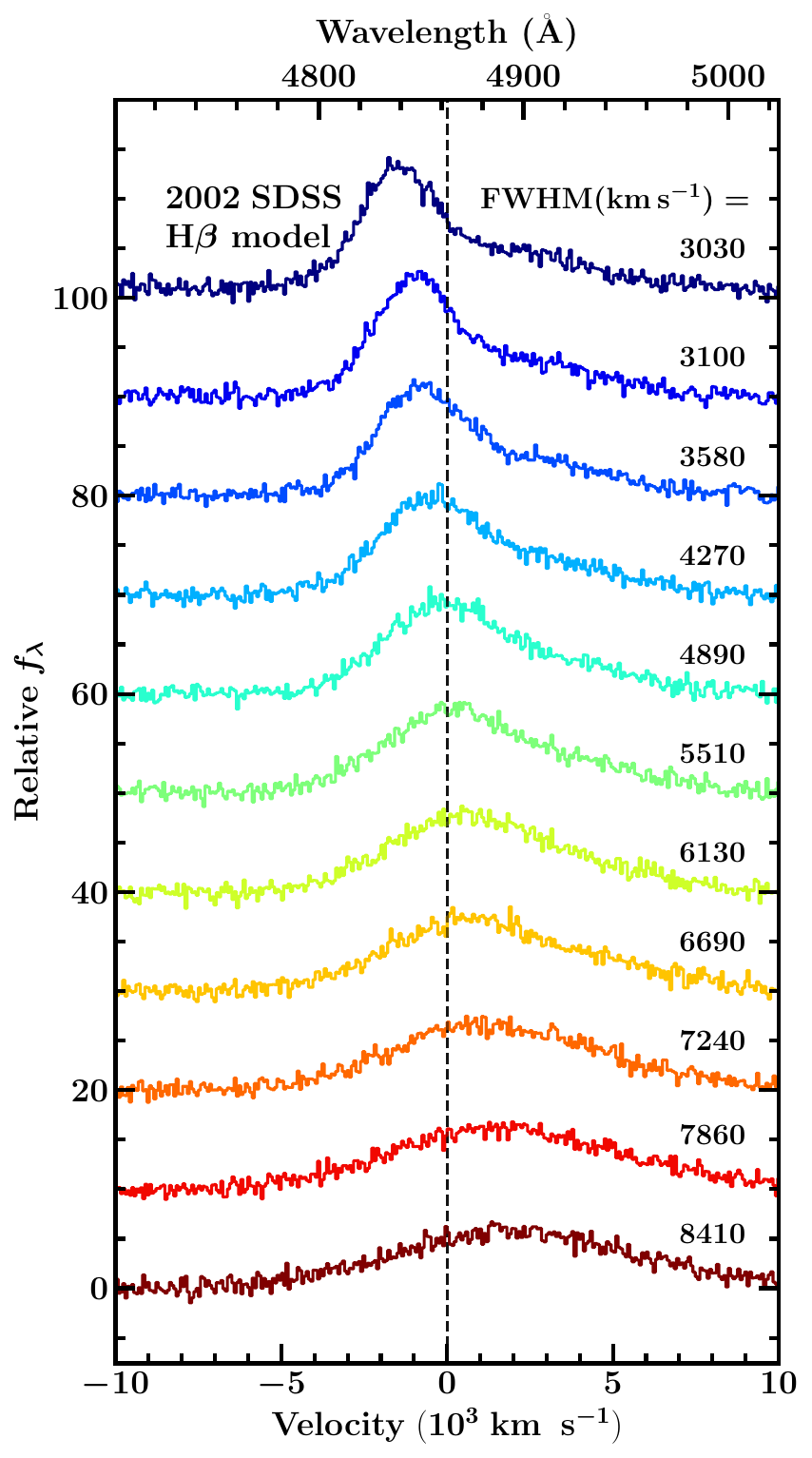}
  \caption{A simulation of the observed variability of the broad H$\beta$ of J0950. The broad H$\beta$ model of the 2002 SDSS spectrum is the basis for producing the other profiles, which have been convolved with Gaussian kernels. These profiles are more round-topped
  than those of the J0950 data (Figure~\ref{fig:decomp spectra}). They serve as a test for the fidelity of our shift measurements for the data––since cross-correlation can reasonably recover the shifts of these simulated profiles, the measurements for the more cuspy broad H$\beta$ lines of the data will only be more accurate (see Figure~\ref{fig:in_out_non_breathing}).}
  \label{fig:sim_non_breathing}
\end{figure}

\begin{figure}
  \includegraphics[width=\columnwidth, angle=0]{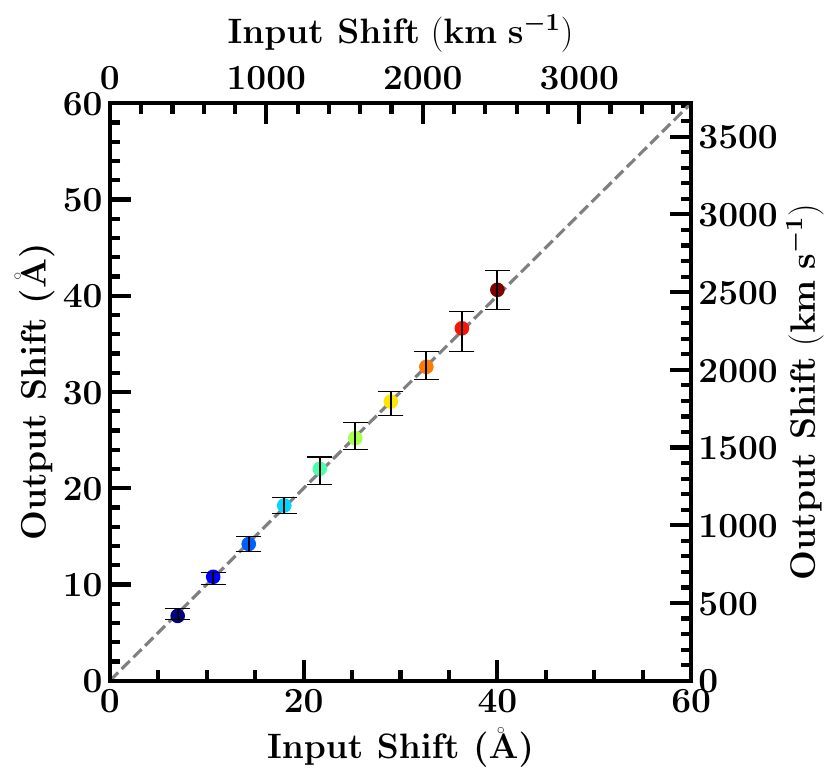}
  \caption{
  Simulation results that demonstrate the reliability of the shift measurements despite the broadening of J0950's H$\beta$ profiles. 
  The observed variability of J0950's broad H$\beta$ profile is simulated (Section~\ref{subsec:non_breathing}; Figure~\ref{fig:sim_non_breathing}) to test if the input shift values can be recovered with cross-correlation. Shown here is
  a plot of
  the input shifts versus the output values, where during cross-correlation, the 2002 SDSS broad H$\beta$ is held constant while each simulated profile is stepped across in wavelength for a $\chi^2$ test.
  The reverse process, where 
  each profile is held constant while the 2002 SDSS broad line is progressively shifted gives the same result. Therefore, it is possible to make reliable shift measurements despite the broadening of the profiles.}
  \label{fig:in_out_non_breathing}
\end{figure}

\subsubsection{ Broad H$\beta$ Velocity Shifts from Parametric Models of the Data} \label{subsec:crosscor_models}
{\color{black}Using the spectral decomposition method described in Section~\ref{subsec:spectral_decomp}, we generated models for the full spectra of J0950. From those, we isolated the broad H$\beta$ models. Models of the full spectra without the contribution from the host galaxy, and isolated broad H$\beta$ lines are shown in Appendix~\ref{appendixA}.}
Following the methodology in Section~\ref{subsec:crosscor_data}, we measured shifts of the broad H$\beta$ using both the full spectra and the isolated broad H$\beta$ emission lines, this time analyzing the models instead of the data.

We first added synthetic Gaussian noise to each spectral model, as demonstrated in Appendix~\ref{appendixA} and then measured the relative shifts of the broad H$\beta$ emission lines by cross correlating the noisy models. 
To quantify the measurement uncertainty, we 
generated 100 realizations of the noise for each model spectrum and repeated the cross-correlation.
We took the median and 68\% confidence interval of the resulting distribution as the final shift measurement and corresponding uncertainty, respectively.

Cross-correlating the broad H$\beta$ lines with the two approaches––using models of the full spectra and using models of the broad H$\beta$ 
profiles only––yielded two sets of relative velocity shifts, shown as {\color{black} a black dotted line and black circles in Figure~\ref{fig:rad_vel_both}}.
The two sets of measurements are qualitatively the same, with the 
velocity shift increasing monotonically in both the full and isolated broad H$\beta$ models. Nevertheless, we performed tests to determine which set of measurements is more reliable. 
Specifically, we cross correlated each isolated broad H$\beta$ line of the data in Figure~\ref{fig:decomp spectra} with its corresponding model {\color{black}(see Appendix~\ref{appendixA}), 
} which, as expected, yielded zero shift.
However, including the Fe~\textsc{ii} lines such that the cross-correlation is between an isolated broad H$\beta$ line from the data and its model of broad H$\beta$ plus Fe~\textsc{ii} lines, gave non-zero shift values. 
Therefore, we conclude that
the presence of the Fe~\textsc{ii} complex introduces a systematic shift in the cross-correlation results.
Moreover, the relative velocity shifts derived from the models ({\color{black}black dotted line and black circles}  
in Figure~\ref{fig:rad_vel_both}) 
allow us to estimate the measurement uncertainty by carrying out multiple realizations of the noise. For these reasons, we adopt the relative velocity measurements based on the model broad H$\beta$ lines throughout this work. 
We also perform the same analysis on the relative velocity measurements from the 
isolated broad H$\beta$ lines of the data, and compare the results (see Section~\ref{subsec:rv_curve}).

{\color{black} 
{\color{black}Furthermore, we used the Fe~\textsc{ii} width experiment described in Section~\ref{subsec:spectral_decomp} to test how the broadening of the Fe~\textsc{ii} template affects the relative velocity shifts obtained by cross correlating the isolated broad H$\beta$ lines. The aim was to assess how uncertainties in the Fe~\textsc{ii} modeling would impact shift measurements. We measured the relative velocity shifts under two scenarios described in Section \ref{subsec:spectral_decomp}: one where the Fe~\textsc{ii} was modeled as narrower, and another where it was modeled as broader than the best-fit width.}
The resulting shifts differ by about 10--75~km~s$^{-1}$, with the larger differences corresponding to the broader H$\beta$ profiles seen in the later spectra of J0950. However, these shifts are all consistent within the measurement uncertainties. The effect of varying the width of the Fe~\textsc{ii} lines is small because the peak of the broad H$\beta$, which is well captured in the modeling, contributes more strongly to cross-correlation measurements than the surrounding, less prominent features. Additionally, the Fe~\textsc{ii} model fits the region under the broad H$\beta$ used for cross-correlation much better than it fits the strong Fe~\textsc{ii} multiplets on either side (see the residuals in Figure~\ref{fig:decomp example}).}
\subsection{Simulating the Profile Change of the Broad H$\beta$} \label{subsec:non_breathing}
The profile of the broad H$\beta$ line of J0950 is observed to change such that it becomes broader over time.  
There is also a $\sim$20\% stochastic variability of the integrated line flux––a small change that is not obvious from a visual inspection of the spectra in Figure~\ref{fig:decomp spectra}; hidden by the fact that the broad H$\beta$ width is increasing.

Profile variability is common in normal quasars, but has been known to produce large uncertainties or even invalidate shift measurements (e.g.,  \citealt{shen2013constraining,liu2014constraining,runnoe2017large}). To test the effects of the observed profile variability of J0950 on our relative shift measurements, we simulated a spectral series based on the model of the broad H$\beta$ line of the 2002 SDSS spectrum. First, we convolved the model with Gaussian kernels of increasing width, and applied velocity shifts to the resulting profiles. 
Gaussian kernels 
offer 
a simple approximation without built-in physical assumptions, 
and allow us to emulate the evolution of the line profile. As we demonstrate in the next paragraph, using Gaussian kernels does not affect our shift measurements. 
Following the procedure in Section~\ref{subsec:crosscor_models}, 
we injected synthetic Gaussian noise into each simulated profile.

We show the simulated spectra in Figure~\ref{fig:sim_non_breathing}, with the broadest spectrum having a full width at half maximum (FWHM) comparable to that of the last 2022 spectrum. 
We then 
carried out the cross-correlation procedure described in Section~\ref{subsec:crosscor_models} 
to find the shift of each simulated profile relative to the 2002 broad H$\beta$ profile. 
The goal of this exercise was to see if cross-correlation can recover the shifts inserted into the simulated profiles. The results are shown in Figure~\ref{fig:in_out_non_breathing}, where each shift value corresponds to the median and  68\% confidence interval of the distribution resulting from 
carrying out multiple realizations of noise before cross-correlation. Figure~\ref{fig:in_out_non_breathing} confirms that the input shifts are reliably 
recovered despite the broadening of the profiles. 
{\color{black} From this simulation we also obtain an estimate of the typical uncertainty in the velocity shifts arising from the increasing width of the H$\beta$ profile with time: 80~km~s$^{-1}$.}
It is important to note that these simulated profiles are more round-topped than the actual observed profiles, all of which appear to have cusps (see Figure~\ref{fig:decomp spectra}). Therefore, if cross-correlation can recover the shifts of profiles that are more round-topped than those of the data, then shift measurements made for the data are also reliable.

\subsection{Simulating Breathing of the Broad-Line Region} \label{subsec:breathing}

\begin{figure}[htbp] 
  \includegraphics[width=\columnwidth, angle=0]{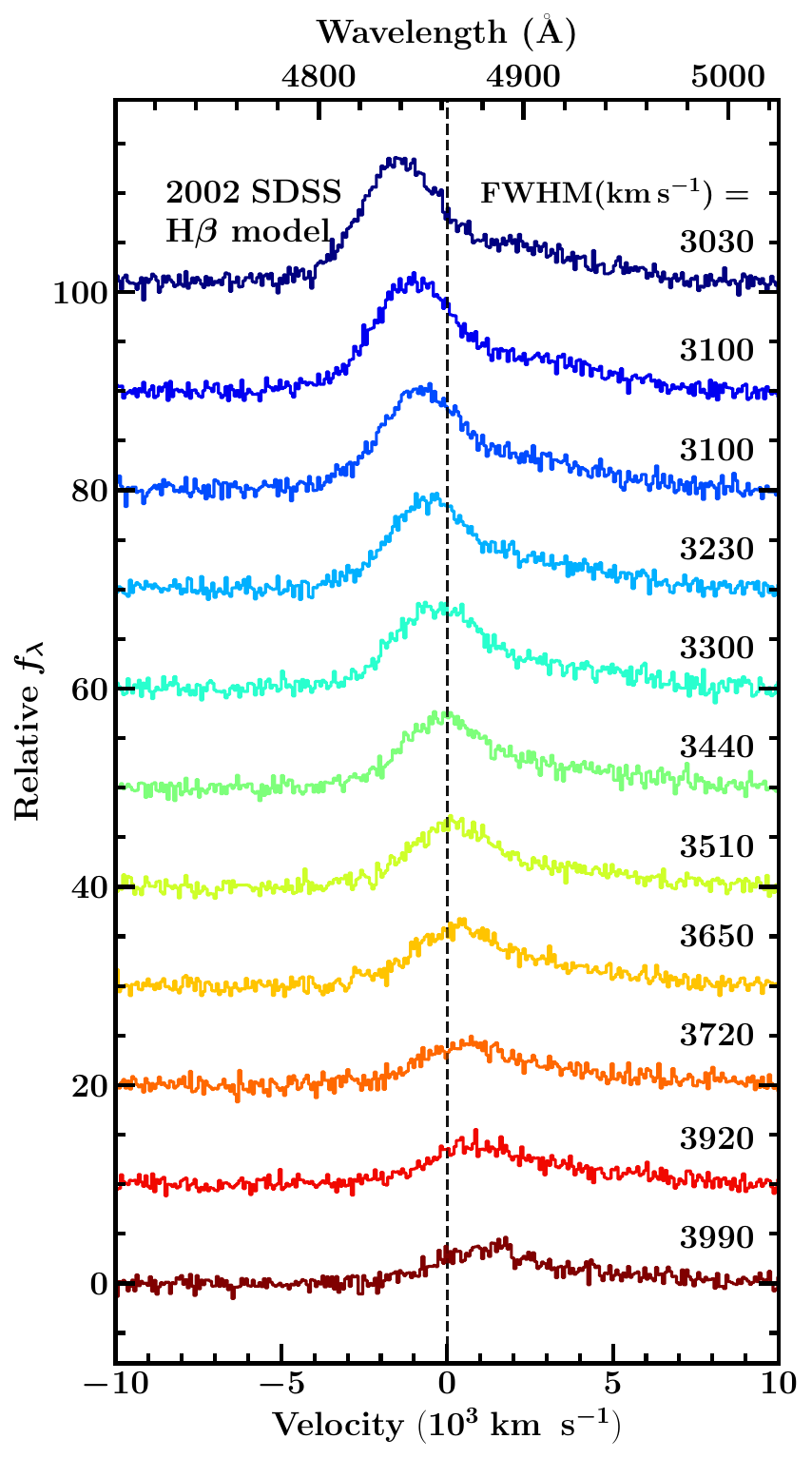}
  \caption{
  Simulation of breathing as a cause of the broad H$\beta$ profile behavior of J0950. The broad H$\beta$ model of the 2002 SDSS spectrum is used as the foundation to construct the remaining profiles. 
  In the simulation, the integrated flux of the broad H$\beta$ line drops faster with FWHM than in the observations (see Figure~\ref{fig:decomp spectra}).}
  \label{fig:sim_breathing}
\end{figure}

\begin{figure}
  \includegraphics[width=\columnwidth, angle=0]{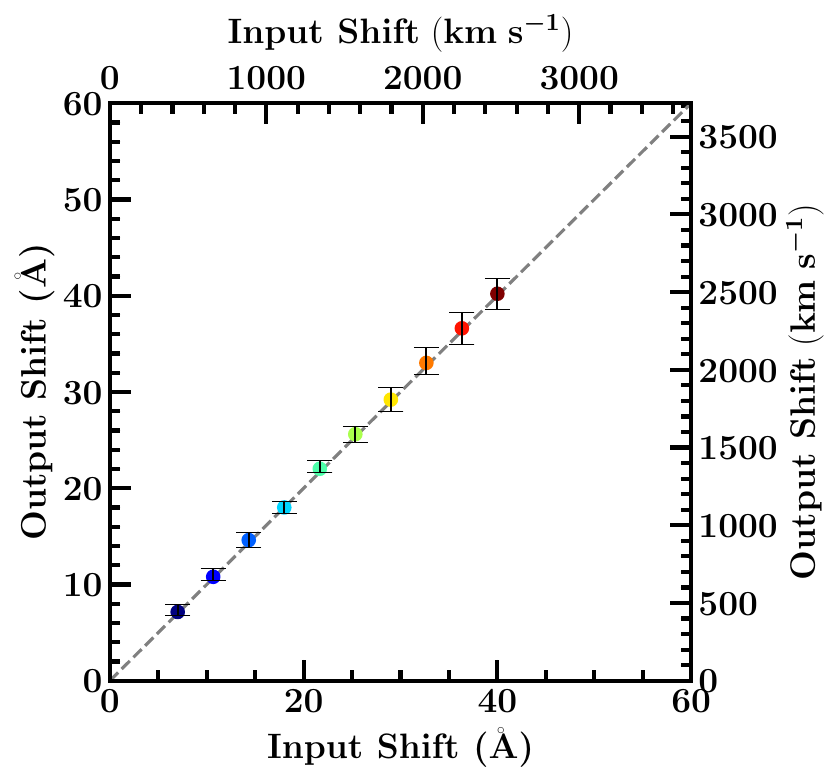}
  \caption{
  The input shift values for the profiles simulated \textit{with breathing} in Figure~\ref{fig:sim_breathing} versus the output values from cross-correlating them,
  where, during cross-correlation,
  the 2002 SDSS broad H$\beta$ is held stationary while each simulated profile is shifted across, and a $\chi^2$ is calculated at each step. 
  The reverse process, where each profile is held constant while the 2002 SDSS broad line is stepped across yields the same result. We see that breathing has negligible effects on shift measurements, since the input values are recovered faithfully.
  }
  \label{fig:in_out_breathing}
\end{figure}

We also explored whether the observed variation of J0950's broad H$\beta$ line can be explained by ``breathing'' of the BLR 
(e.g., \citealt{peterson2011masses,wang2020sloan}). Breathing 
is the effect in which changes in the accretion-powered continuum of a quasar affect the extent of the zone in the BLR that can emit a line efficiently, hence the widths of the resulting broad lines become broader as those lines become fainter.
To examine if the broad H$\beta$ behavior of J0950 can be replicated by breathing, and if that has an impact on our shift measurements, we repeated the exercise of Section~\ref{subsec:non_breathing} with the difference that we simulated the spectra by applying breathing effects. Specifically, we used a relationship derived under the premise that the 
observed continuum flux $F$ and width of the broad H$\beta$ line are connected to the mass of the BH by 
$M_{\rm BH}\propto F^{1/2}\cdot (\mathrm{FWHM})^2$ (\citealt{bentz2009radius,peterson2011masses}). 
Since $M_{\rm BH}$ is constant, we can write


\begin{equation}\label{reverb2}
\begin{aligned}
F_{\rm after}= F_{\rm before}\Big(\dfrac{\mathrm{FWHM}\,_{\rm before}}{\mathrm{FWHM}\,_{\rm after}}\Big)^4
\end{aligned}
\end{equation}
where $F_{\rm before}$ and $\mathrm{FWHM}\,_{\rm before}$ are the integrated flux and width of the broad H$\beta$ emission respectively. $F_{\rm after}$ and $\mathrm{FWHM}\,_{\rm after}$ are the flux and width of the profile resulting from the convolution of the broad H$\beta$ line with a wider Gaussian kernel. 
Figure~\ref{fig:sim_breathing} shows the sequence of spectra that simulate symmetric profile variability expected from BLR breathing. The shift values recovered from cross-correlation are shown in Figure~\ref{fig:in_out_breathing}. 

This exercise leads us to conclude the following: (a) the decline in integrated flux with increasing width of the broad lines in this simulation (Figure~\ref{fig:sim_breathing}) is more drastic than that observed in the broad H$\beta$ line of J0950 (compare Figure~\ref{fig:decomp spectra}). 
Moreover, as noted at the beginning of Section~\ref{subsec:non_breathing}, the broad H$\beta$ flux of this object exhibits only a $\sim$20\% stochastic variability
––a behavior inconsistent with breathing since the FWHM increases systematically by a factor of $\sim3$ over the monitoring period. 
(b) It is clear from Figure~\ref{fig:in_out_breathing} that known shifts from the simulation can be measured despite the effects of breathing, which leads us to infer that symmetric profile variability, in general, has a negligible effect on the shift measurements. 

\subsection{The Radial Velocity Curve} \label{subsec:rv_curve}

In Sections \ref{subsec:crosscor_data}$-$\ref{subsec:crosscor_models} we obtained four different sets of relative velocity shifts by cross-correlating the broad H$\beta$ lines in: (a) the full spectra of the data, (b) the isolated broad H$\beta$ lines of the data, (c) the full spectra models, and (d) the isolated broad H$\beta$ models. In Section~\ref{subsec:crosscor_models}, we established that the presence of Fe~\textsc{ii} lines in the full spectra (both data and models), introduces a systematic shift in the cross-correlation results, making shift measurements from the full spectra less reliable. 

In view of the above, we proceeded only with the relative shift measurements based on the isolated broad H$\beta$ lines 
(Figure~\ref{fig:decomp spectra}), and their corresponding models (Appendix~\ref{appendixA}). We added to these shift values the peak offset measurement of the broad H$\beta$ of the 2002 spectrum. To measure the peak offset (following a method similar to \citealt{eracleous2012large}), we defined a fixed region around the peak and repeatedly fitted it with a Gaussian, each time varying the wavelength range used for fitting within that region. We took the median peak velocity as the best estimate of the true value and the 68\% confidence interval of the resulting distribution of peak values as the measurement uncertainty. It is important to note that this method of locating the peak is sensitive to asymmetries within the broad emission line. Therefore, the Gaussian fitting region must be restricted to be near the peak, excluding significant asymmetries. 

The addition of the peak offset from the 2002 spectrum provides the absolute peak offsets of the broad lines, 
as listed in Table~\ref{tab:rel_shifts}.
{\color{black}The measurements agree within 3$\sigma$, with average differences around 2$\sigma$.}
{\color{black}We return to these measurements 
in Section~\ref{subsec:rv_fitting} to discuss the implications.}


\begin{deluxetable*}{lccccccc} 
\tablewidth{0pt} 
\tablecaption{Absolute peak offset measurements 
for isolated broad H$\beta$ lines of the data, and model broad H$\beta$ lines
\label{tab:rel_shifts}}
\renewcommand{\arraystretch}{1.2} 
\tablehead{
\colhead{Observation} & \colhead{Data H$\beta$} & ~ & \colhead{Model H$\beta$} & ~ 
\\
\colhead{Date (UT)} & \colhead{(km s$^{-1}$) $^a$} & 95\% $^b$ & \colhead{(km s$^{-1}$) $^c$} & 95\% $^b$
}
\startdata
2002.05.15 & $-1360^{+60}_{-50}\,^d$ & ($^{+110}_{-140}$) & $-1360^{+60}_{-50}\,^d$ & ($^{+110}_{-140}$) 
\\
2010.05.08 & $-920^{+40}_{-40}$ & ($^{+100}_{-90}$) & $-890^{+60}_{-40}$ & ($^{+100}_{-100}$) 
\\
2011.04.08 & $-660^{+40}_{-40}$ & ($^{+90}_{-90}$) & $-660^{+60}_{-50}$ & ($^{+110}_{-100}$) 
\\
2015.01.20 & $-250^{+40}_{-40}$ & ($^{+90}_{-90}$) & $-180^{+50}_{-40}$ & ($^{+100}_{-100}$) 
\\
2019.04.09 & $560^{+40}_{-40}$ & ($^{+90}_{-90}$) & $460^{+50}_{-50}$ & ($^{+120}_{-90}$) 
\\
2020.02.02 & $730^{+40}_{-40}$ & ($^{+90}_{-90}$) & $790^{+70}_{-80}$ & ($^{+140}_{-160}$) 
\\
2020.04.02 & $720^{+40}_{-40}$ & ($^{+90}_{-90}$) & $610^{+80}_{-80}$ & ($^{+240}_{-180}$) 
\\
2021.04.05 & $870^{+40}_{-40}$ & ($^{+90}_{-90}$) & $760^{+210}_{-90}$ & ($^{+340}_{-180}$) 
\\
2022.03.25 & $1440^{+40}_{-60}$ & ($^{+110}_{-120}$) & $1160^{+110}_{-80}$ & ($^{+260}_{-130}$) 
\\
2022.04.03 & $1560^{+40}_{-40}$ & ($^{+90}_{-90}$) & $1330^{+170}_{-140}$ & ($^{+300}_{-230}$) 
\\
2022.12.31 & $1760^{+40}_{-40}$ & ($^{+90}_{-90}$) & $1510^{+110}_{-120}$ & ($^{+190}_{-280}$) 
\\
2024.03.30 & $2350^{+40}_{-40}$ & ($^{+90}_{-90}$) & $2140^{+110}_{-90}$ & ($^{+210}_{-160}$) 
\\
\enddata
\tablecomments{Section~\ref{subsec:rv_curve} describes how the absolute shifts were obtained. {\color{black}In addition to the tabulated uncertainties, there is a systematic uncertainty of $60$~km~s$^{-1}$ from Fe~\textsc{ii} subtraction (Section \ref{subsec:crosscor_models}), a systematic uncertainty of $80$~km~s$^{-1}$ due to the increase in the width of the broad H$\beta$ profile with time (Section \ref{subsec:non_breathing}), and a 200~km~s$^{-1}$ jitter (Section~\ref{subsec:rv_fitting}).}\\
$^a$ Absolute shifts and {\color{black}68\% confidence intervals} of the isolated, broad H$\beta$ lines of the data shown in Figure~\ref{fig:decomp spectra}. 
\\
     $^b$ {\color{black}95\% confidence levels for the absolute shifts.}
\\
    $^c$ Absolute shifts and {\color{black}68\% confidence intervals} of the isolated, model broad H$\beta$ shown in Appendix~\ref{appendixA}.
\\
    $^d$ H$\beta$ peak offset of the 2002 SDSS spectrum.
    } 
    
\end{deluxetable*}

\subsection{A Summary of Uncertainties}\label{subsec:uncertainties}
{\color{black}Throughout Section~\ref{sec:methods}, we measured velocity shifts of the broad H$\beta$ line and did a series of tests and simulations to characterize the uncertainties in those measurements. The formal uncertainties from cross correlation are described in Section~\ref{subsec:crosscor} and are listed in Table~\ref{tab:rel_shifts}; they are between 40 and 60~km~s$^{-1}$. A systematic uncertainty of about 60~km~s$^{-1}$ arises from the modeling and subtraction of the Fe~\textsc{ii} complex (Section~\ref{subsec:crosscor_models}). From the simulation in Section~\ref{subsec:non_breathing}, we estimated an uncertainty of $\sim$80~km~s$^{-1}$ due to the increasing width of the broad H$\beta$ line over time. To account for short-timescale radial velocity fluctuations, such as the example in Section~\ref{subsec:crosscor_data}), we adopted a jitter of 200~km~s$^{-1}$ (discussed further in Section~\ref{subsec:rv_fitting}) for use in the radial velocity analyses presented later. The jitter is the dominant source of uncertainty.}

\section{Analysis of Light Curves}\label{sec:lightcurves}
\begin{figure*}
    \centering
    \includegraphics[width=6.5in]{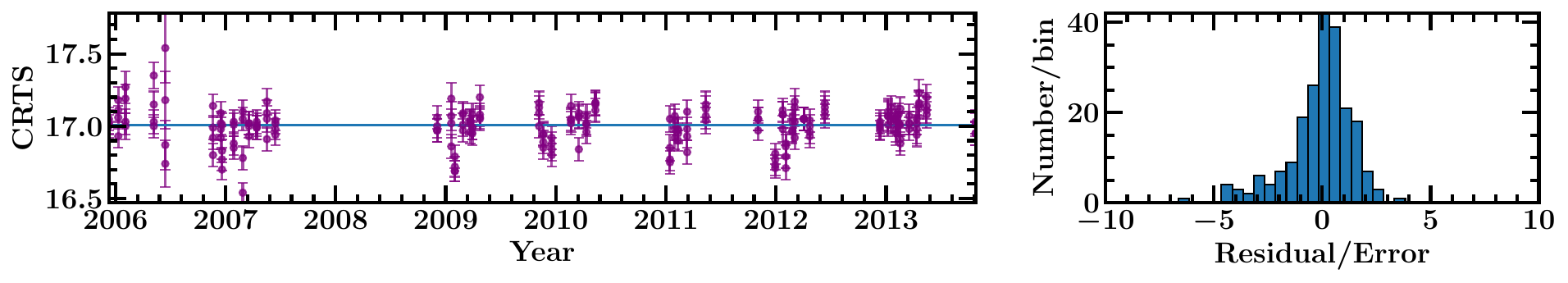} \\
    \includegraphics[width=6.5in]{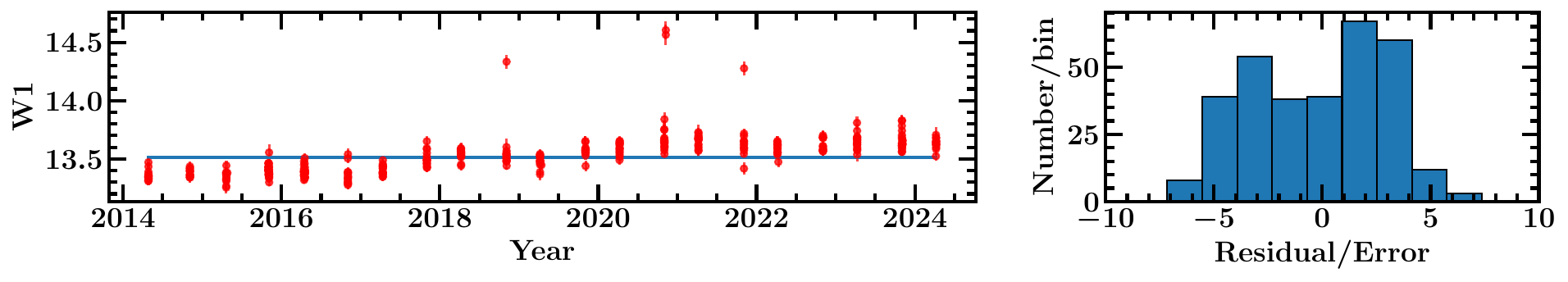} \\
    \includegraphics[width=6.5in]{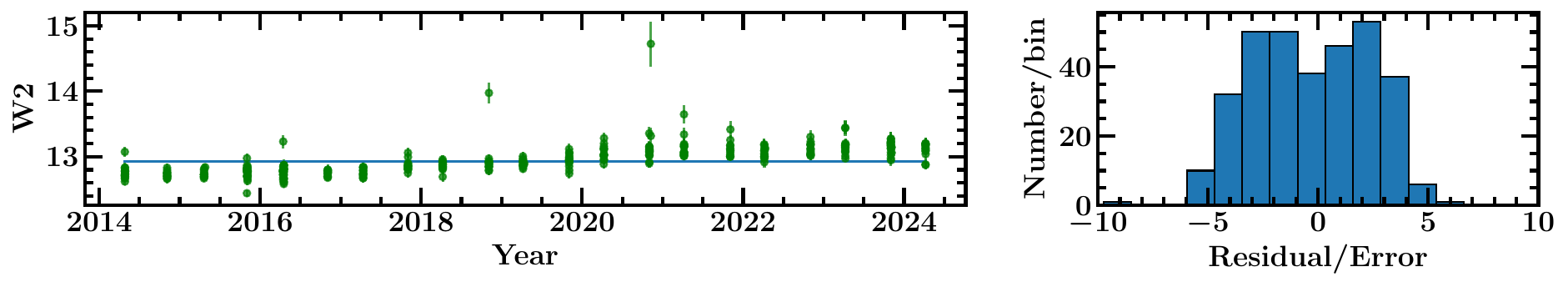} \\
    \includegraphics[width=6.5in]{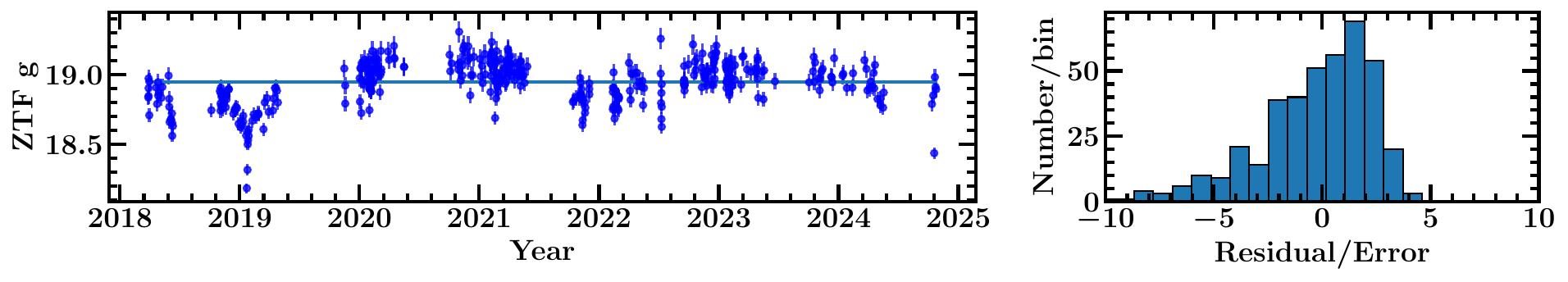} \\
    \includegraphics[width=6.5in]{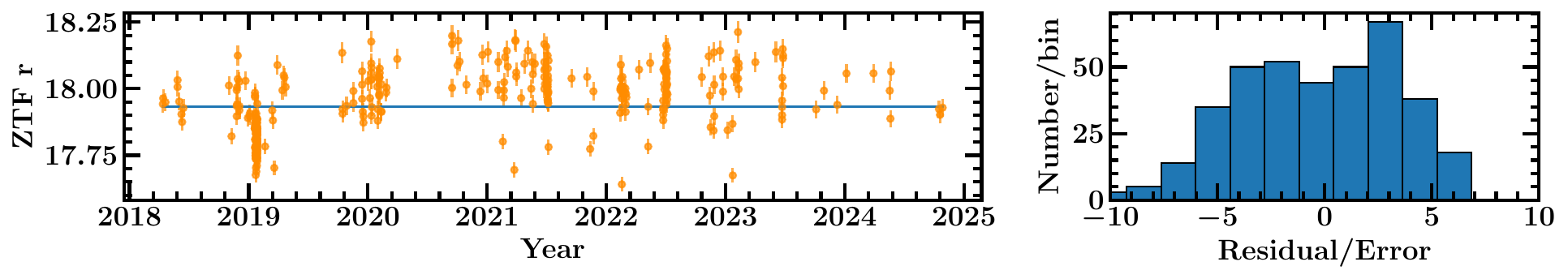} \\
    \includegraphics[width=6.5in]{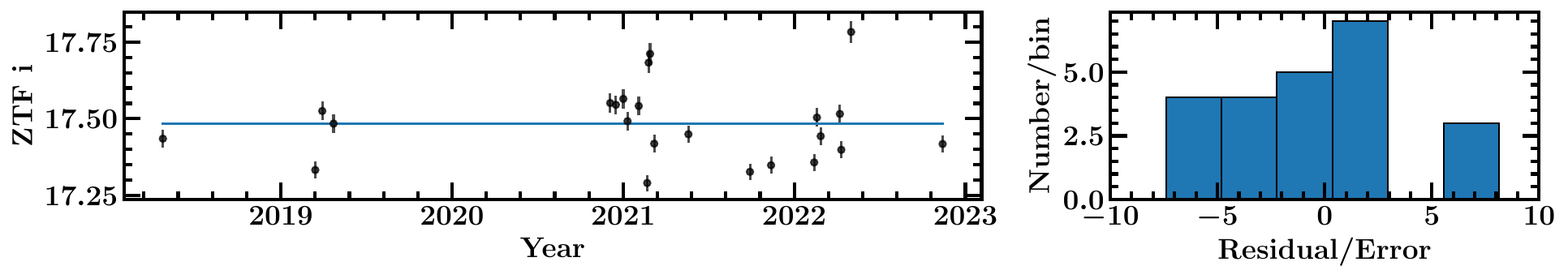}
\caption{\color{black} \textit{Left}: The light curves of J0950 from CRTS, NEOWISE, and ZTF in different bands, each fitted with a constant. \textit{Right}: Distributions of the residuals normalized by the measurement uncertainties of the light curves on the \textit{left}. The scatter in the distributions indicate some intrinsic variability, but not a strong deviation from a linear trend.
}
\label{fig:residuals}
\end{figure*}

{\color{black}
We investigated the optical and infrared (IR) light curves of J0950 using all publicly available data from the Catalina Real-Time Transient Survey (CRTS), the Near-Earth Object Wide-Field Infrared Survey Explorer (NEOWISE), and the Zwicky Transient Facility (ZTF). Although the available spectroscopic data span a longer time period than these light curves, we nonetheless analyzed the full extent of the available photometric data, shown in Figure~\ref{fig:light_curve}.

While there is no apparent periodicity in the light curves, we computed the Lomb-Scargle Periodogram\footnote{We utilized the \texttt{LombScargle} feature from the \texttt{astropy.timeseries}  \textit{python} package.} for the data from each facility separately, as well as for the combined light curve, with each dataset normalized to the CRTS light curve by its median magnitude.  
The period search was limited to periods up to 60\% of the total time span of each dataset. Periods longer than the dataset duration are undetectable, and those longer than half the time span are unreliable, as they would not allow for the observation of at least two full cycles. 
In the periodograms, we set significance levels corresponding to 0.1\% and 1\% false alarm probabilities (corresponding to 99.9\% and 99\% confidence respectively), ensuring that any meaningful signal stands significantly above both the noise floor and these thresholds. The only significant signals appear at short periods corresponding to the observational cadences, around 1~yr.

In addition to the periodograms, we modeled each light curve with a flat line corresponding to its 3$\sigma$-clipped mean magnitude, and computed residuals normalized by the measurement uncertainties, as shown in Figure~\ref{fig:residuals}. The distributions of residuals normalized by the error bars 
appear approximately symmetric, though with a broader spread than 
a standard normal distribution, likely due to intrinsic stochastic variability of the quasar \cite[e.g.,][]{vaughan2016false, liu2018did}. {\color{black}The fit of a flat line in Figure~\ref{fig:residuals} and the lack of any any periodic signal in the periodograms, leads us to conclude that there is no evidence of periodicity in the light curves.} 
The light curves of J0950 are revisited in Section~\ref{subsec:doppler_boost}.}

\section{Interpretation, Discussion, and Speculation}\label{sec:conclusion}
\subsection{Summary of Main Results from the Data}

The quasar J0950 was suggested as a spectroscopic SBHB candidate 
based on  the large velocity shift of its broad H$\beta$ line.
Subsequent observations of this object reveal a progressive and systematic shift of its broad H$\beta$ line over time. It is particularly noteworthy that the velocity of the line changed sign during its evolution.
The profile shape of the broad H$\beta$ line also changed between observations.

We used a cross-correlation technique to measure the velocity shifts of the broad H$\beta$ lines, applying it to both the observed line profiles 
and corresponding models. In each case, we used the full spectra of J0950 as well as the isolated H$\beta$ emission lines obtained through spectral decomposition. The relative velocity shifts from all approaches show a monotonic increase from negative to positive velocities. After extensive testing, we adopted the relative velocities 
derived from cross-correlating models of the isolated broad H$\beta$ lines as the most reliable.
We also carried out the same analysis on the isolated broad H$\beta$ lines of the data in order to quantify systematic uncertainties.
The choice of the isolated broad H$\beta$ models 
mitigates the systematic errors introduced by Fe~\textsc{ii} lines, {\color{black}and the velocity measurements from these models provide an estimate of the measurement uncertainties through multiple noise realizations.} 

In addition to displaying a velocity shift, the broad H$\beta$ line profile of J0950 gradually broadened over time. {\color{black}This increase in width does not compromise our ability to measure shifts but introduces an additional uncertainty comparable in magnitude to the cross-correlation measurement uncertainties. The observed broadening is not consistent with the ``breathing'' behavior of a broad-line region.}

{\color{black}Finally, we analyzed available light curves from CRTS, NEOWISE (W1 and W2 filters), and ZTF (\textit{g}, \textit{r}, and \textit{i} bands). The light curves display a low level of apparently stochastic variability with no evidence for periodicity.}


\subsection{Assessment of Different Interpretations for the Observed Velocity Shifts}\label{subsec:assess_interp}

{\color{black}
In this section, we consider the interpretations listed in Section~\ref{subsec:possible_interp} for the observed shift of the broad H$\beta$ line of J0950. We discuss those interpretations and evaluate whether they are physically plausible and applicable, before contemplating whether they can be subjected to statistical tests.} 

\begin{enumerate}[label=(\alph*)]
{\color{black}
\item
The broad H$\beta$ velocity shifts in the spectra of J0950 have increased steadily over the decades it has been monitored (Section~\ref{subsec:rv_curve}). Such a trend resembles closely the expected signature of a long-period (several decades) binary supermassive black hole with a BLR associated with one of the two objects. The shifts in the broad H$\beta$ are then associated with the changes of the projected bulk velocity of the BLR over a portion of an orbital cycle. Since the time scale and general pattern of the radial velocity variations is consistent with expectations \citep[e.g.,][]{eracleous2012large,runnoe2017large}, we explore this hypothesis further by fitting a Keplerian orbit model to the observed radial velocity curve in Section~\ref{subsec:rv_fitting} and considering the implications in Sections~\ref{subsec:explore_sbhb} and \ref{subsec:doppler_boost}.

\item
As a dust cloud outside the BLR and in a Keplerian orbit crosses our line of sight, it can obscure one side of the BLR first, then block a significant portion of the innermost BLR before obscuring the other side. Once the cloud moves out of our line of sight to the BLR, the full region becomes visible again. In a spectroscopic time series, this would result in the alternating appearance of the blueshifted and redshifted sides of a broad emission line. 
When the cloud is blocking the innermost BLR, the observed broad emission line appears centered, narrow and has a small integrated flux, since only the low-velocity gas in the outer parts of the BLR is visible. When the cloud is out of our line of sight to the BLR, a much stronger, broader line profile 
is observed because the full velocity field is visible.
Therefore, a moving dust cloud can produce asymmetric broad emission lines and create apparent velocity shifts by selectively obscuring parts of the velocity field. However, the evolution of J0950's broad H$\beta$ line is not consistent with this scenario. While we do observe blueshifted and redshifted broad-line peaks in the data, the dust-obscuration scenario is unlikely, since the spectrum in which the broad H$\beta$ line appears centered and relatively symmetric (2015 SDSS spectrum; Figure~\ref{fig:data}) is not extra strong and broad, as would be expected if the dust cloud were out of our line of sight to the BLR, nor narrow and weak, as would be expected if dust were temporarily obscuring the innermost parts of the BLR. Moreover, the observed variations of the broad H$\beta$ flux are only 20\%. 

We can also estimate the crossing time for an obscuring object orbiting the BLR by following the method of \cite{lamassa2015}. For a $10^8 \,\mathrm M_{\odot}$ supermassive BH with a BLR radius of $0.15$~pc 
(see estimates in Section~\ref{subsec:explore_sbhb}), it would take $\sim260$~years for a dust cloud in Keplerian orbit at that radius to cross the BLR. Given that the broad H$\beta$ line of J0950 exhibits a significant blueshift and redshift in the span of $\sim22$~years, this timescale discrepancy makes it difficult to reconcile the observations with the dust-obscuration scenario. Furthermore, matching the observed timescale of about 20~years would require a BH mass $\gtrsim 10^{10} \,\mathrm M_{\odot}$, which is physically implausible \citep[e.g.,][]{thomas201617,inayoshi2016there,mcconnell2012dynamical}.

\item
Another possible source of shifted broad emission lines is outflows. In the case of bipolar outflows \citep[e.g.,][]{zheng1990outflows,veilleux1991outflows,zheng1991outflows} 
where the ejection is symmetric and expands outwards, the observer would first see the blueshifted side, followed by the redshifted side after a time delay. In this scenario, both the red and blue sides of a broad emission line must be visible. The only way for the blue side to appear first, then disappear, and then the red side appear, as we observe in J0950, would be if the blue side were to turn off exactly as the red side turns on. This would require exquisite fine tuning. Even with such fine tuning, the bipolar outflow scenario cannot explain the symmetric 2015 line profile centered at $v=0$.

\item
The possibility that J0950 is a recoiling BH is unlikely on several grounds. A recoiling BH, after being ejected, could reach a maximum distance from the galactic center up to several kpc \citep[e.g.,][]{blecha2016recoil} before falling back and oscillating around the nucleus. However, the timescale for one full oscillation cycle is at least $10^6$~years \citep{komossa&meritt2008recoil}, which is far longer than the duration over which J0950 has been observed. 
Furthermore, although recoil speeds in excess of several thousand $\mathrm{km\,s^{-1}}$ are possible in principle \citep[e.g.,][]{healy2023recoil}, the probability of observing a kick velocity that is $\gtrsim1000 \, \mathrm{km\,s^{-1}}$ from a supermassive binary merger is negligibly small; most recoil velocities are on the order of a few hundred $\mathrm{km\,s^{-1}}$ or less \citep{dotti2010dual}.\footnote{Recoils from three-body interactions, can reach $\sim$1000 km~s$^{-1}$ for the lightest supermassive BH in a three-body interaction but are typically on the order of a few hundred $\mathrm{km\,s^{-1}}$ \citep{hoffman2006recoil}.}
The likelihood of detecting kick velocities much greater than the escape speed of the host galaxy is even lower. In the case of J0950, where we observe velocities as high as $\sim2000\,\mathrm{km\,s^{-1}}$, that probability is 
negligibly small \citep{blecha2016recoil}. Most critically, the radial velocity curve of J0950 mostly shows consistent positive acceleration as well as a change in sign––a behavior that is inconsistent with the expected dynamics of a recoiling BH. 

}

\item
There are known objects that show a single displaced peak with a shoulder, similar to the earliest spectrum of J0950. Examples include Mrk 668 and 3C 227, both of which have broad emission lines with a displaced peak and shoulder (\citealt{marziani1993peculiar, gezari2007long}). They have been monitored, more sparsely than J0950, and some do show radial velocity variations, but none have the pattern we see in J0950 (especially not the change of sign of the radial velocity curve). Non-axisymmetric disk models (e.g., \citealt{storchi2017double,schimoia2017evolution}) have been used to try to explain their irregular, non-periodic radial velocity curves. 
{\color{black}These models entail a ``perturbation'', a bright spot or spiral arm, orbiting in the inner region of a circular, disk-like BLR, or a precessing elliptical disk \citep{gezari2007long,lewis2010long,storchi2017double}.}
This scenario can cause a broad emission line with two peaks, whose strengths alternate like a seesaw as the perturbation moves and enhances one peak after another. {\color{black}The smoothly increasing radial velocity variation of J0950 differs} from the variations of these objects and cannot be explained by the non-axisymmetric BLR models described above. {\color{black}Nonetheless, the broader framework of a perturbed BLR may provide a plausible explanation for J0950's behavior. Since other objects within this framework have shown substantial changes in their radial velocity over time, continued monitoring of J0950 is essential to see whether its trend persists or evolves (see the discussion of 3C 390.3 in Section 4.2 of \citealt{eracleous1997rejection}).} 
{\color{black}Further testing of the single perturbed BLR scenario for J0950 will require developing a detailed physical model, which is out of the scope for this paper.}

{\color{black}J0950 is an unusual object. Among quasars, only double-peaked emitters show behavior even remotely resembling the large monotonic velocity variations seen in J0950. \cite{doan2020improved} present the longest monitoring campaign for double-peaked emitters; only two objects (3C~332 and 3C~390.3) displayed fairly monotonic velocity changes over most of their monitoring periods. Since double-peaked emitters make up only $\sim$3\% of all quasars \citep{strateva2003doublepeaked}, such large monotonic variations are extremely rare––roughly 0.4\% of quasars (2/14 of extensively monitored double-peaked emitters $\times$3\%). Importantly, none of these objects show a change in sign in their radial velocity curves. Therefore, the velocity changes observed in J0950 cannot be directly attributed to phenomena seen in other quasars.}

\end{enumerate}

{\color{black}Following the discussion above, we now explore in more detail whether the binary supermassive black hole interpretation is consistent with the available data. As noted earlier we will examine the single, perturbed BLR interpretation in detail in the future. 
We consider the remaining scenarios implausible and do not discuss them further.}

\subsection{A Supermassive Binary Black Hole Scenario for the Radial Velocity Curve}\label{subsec:rv_fitting}

\begin{figure*}
  \centering
  \includegraphics[scale=0.55, angle=0]{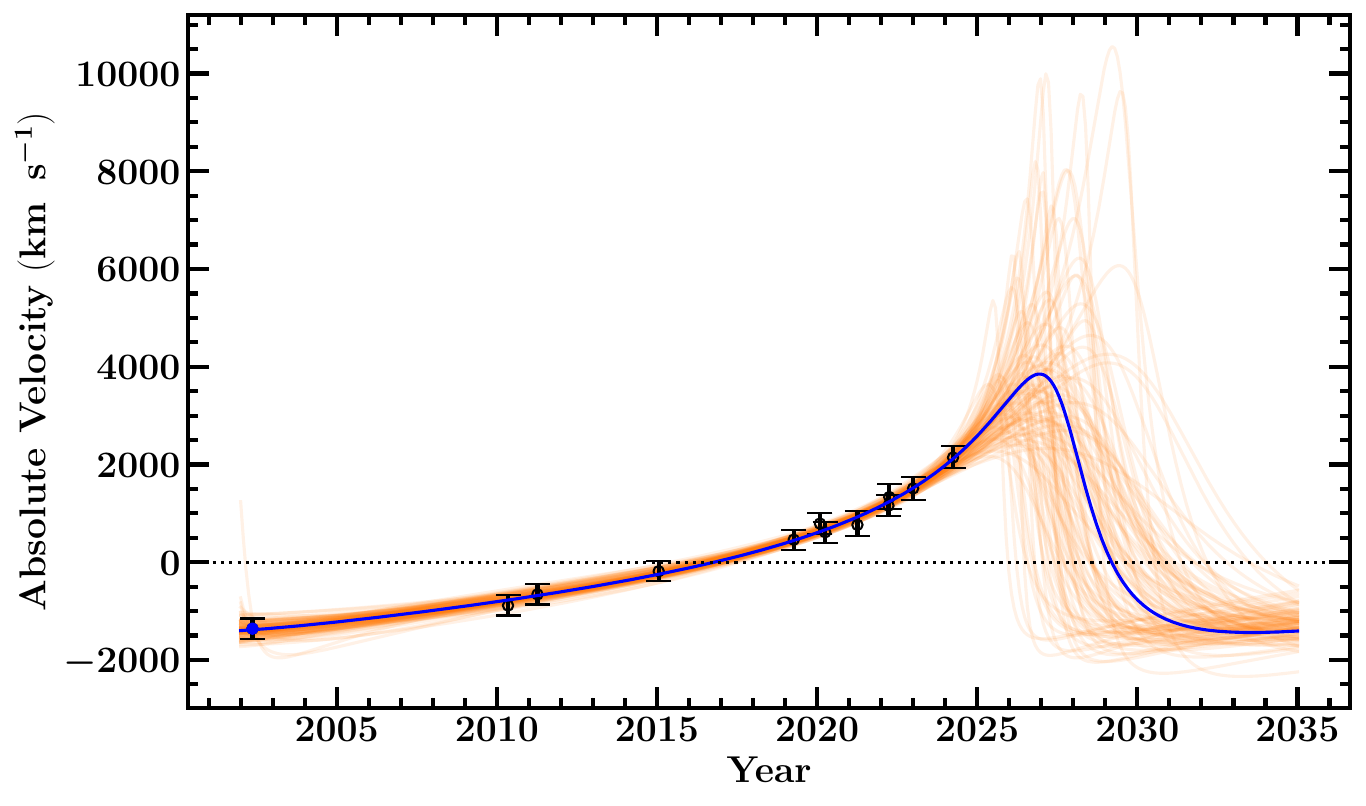} 
  \caption{
Radial velocity curve of J0950 based on the model broad H$\beta$ lines. The rest wavelength of H$\beta$ corresponds to the horizontal dotted line at 0 km s$^{-1}$, relative to which the velocity offsets are measured. The error bars correspond to 68\% confidence intervals with jitter of 200 km s$^{-1}$ included in quadrature. The first point marks the peak offset of the broad H$\beta$ of the 2002 spectrum, relative to which all other offsets are measured.
The thick blue line represents the best-fitting eccentric orbit model, with corresponding orbital parameters given in Table~\ref{tab:model_params}, while the light orange lines show 100 other solutions drawn from the posterior distribution. 
}
  \label{fig:rad vel}
\end{figure*}

{\color{black}The radial velocity curve of J0950 based on the fourth column of Table~\ref{tab:rel_shifts} is shown in Figure~\ref{fig:rad vel}. It shows a steady increase of $\sim$3500~km~s$^{-1}$ over time; the most striking behavior 
is that the broad line region of this quasar has} undergone radial velocity variations in a manner resembling Keplerian motion in a binary. 
There certainly are potential caveats, which we discuss in Section~\ref{subsec:assess_interp}. But the radial velocity curve of J0950 so far is consistent with binary motion, which makes it a strong SBHB candidate.

We increased 
the {\color{black}measurement uncertainties of the radial velocity curve} to capture the effects of 
``jitter'' from regular quasar variability (\citealt{guo2019constraining, doan2020improved}). 
Quasars can have perturbations or asymmetries within the broad line region that manifest as small, stochastic radial velocity variations, or jitter, occurring over a 
timescale of the order of a year or less (e.g., \citealt{barth2015lick}). 
{\color{black}This subtle effect is directly measurable in two of J0950's spectra (see Section~\ref{subsec:crosscor_data}).}
\cite{doan2020improved} empirically determined that jitter can be characterized as fluctuations around a smooth radial velocity curve, distributed approximately according to a Gaussian distribution (see also \citealt{guo2019constraining}). The standard deviation of the distribution is on the scale of a few hundred km~s$^{-1}$. 
Following the methodology from that work, we added a jitter of 200~km s$^{-1}$ in quadrature to the uncertainties in each radial velocity measurement. {\color{black}
The jitter dominates over the measurement errors and is the primary source of uncertainty in the radial velocity at each epoch.}

{\color{black}To fit the radial velocity curve shown in Figure~\ref{fig:rad vel}, we used \texttt{radvel} (\citealt{fulton2018radvel}), a package for fitting Keplerian orbits to radial velocity time series.}  
The best-fitting orbital parameters, determined from the radial velocity curve fit {\color{black}in the observed frame,} are provided in Table~\ref{tab:model_params}. 
The free parameters in the fitting are the period ($P$), the time of conjunction ($T_c$), the natural log of the semi-amplitude (ln $K$), and $\sqrt{e}\,\mathrm{cos}\,\omega$ and $\sqrt{e}\,\mathrm{sin}\,\omega$, from which the eccentricity ($e$) and argument of periapsis ($\omega$) are derived.
{\color{black}Given the exploratory nature of fitting the velocity curve of J0950 as a binary and the absence of prior knowledge about the system, we adopted broad, physically motivated priors based on 
the radial velocity curve. {\color{black}Specifically, we assumed a near-circular orbit as a starting point and adopted Gaussian priors on the period and time of conjunction: a mean period of $40\pm15$~yr (ensuring that the 1-$\sigma$ range does not fall below the 22~yr span of radial velocity measurements available), and a time of conjunction $2016\pm20$~yr. For a circular orbit, the conjunction––the moment when the lower-mass, or secondary BH, passes along the line of sight between the observer and the primary BH––would be expected near that epoch. 
A hard bound of $6<{\rm{ln}}\,K<9$ was applied to the semi-amplitude to restrict unphysical amplitudes, and an eccentricity prior of $e<1$ was imposed.} 
The Markov chain Monte Carlo (MCMC) {\color{black}was run with 20 walkers for $\sim$2500 steps each}, and the chains converged reliably after $\sim 3\times10^5$ steps (Gelman-Rubin statistic $<$~1.01) {\color{black}following an initial burn-in}. The resulting posteriors were well-constrained and consistent with the observed signal (see Appendix~\ref{appendix_corner} for a corner plot), indicating that the priors were non-restrictive. 
}

\bluetext{Since the radial velocity baseline of 22~yr covers only a fraction of the 33-yr orbit, the period, time of conjunction and the semi-amplitude, become partially degenerate. Therefore, the posterior distribution at longer periods is strongly influenced by the assumed prior. For much longer periods, the radial velocity curve would look almost the same over the observed 22~yr window. Likewise, the semi-amplitude can shift slightly depending on the adopted jitter value. These effects are expected for long period systems.} \mgtext{In order to improve the sensitivity of the data to longer orbital periods, further monitoring of the radial velocity curve of J0950 is needed for a substantial amount of time.}

{\color{black}The median values of the posterior distributions derived from the MCMC analysis are reported in Table~\ref{tab:model_params}.} 
The uncertainties are 68\% confidence limits.
Table~\ref{tab:model_params} also provides the orbital parameters from performing the same analysis 
on the radial velocity curve obtained from the absolute shifts of the isolated, broad H$\beta$ lines of the data {\color{black}(provided in the second column of Table~\ref{tab:rel_shifts})}. 
The results of the two different fits capture the magnitude of the systematic uncertainties.  
The uncertainties encompass a wide 
range of possible Keplerian orbit models, such as those illustrated by the light orange lines in Figure~\ref{fig:rad vel}. Observing a turnover in the radial velocity curve within the next few years would significantly constrain the possible models. 

\begin{deluxetable*}{lcc}[p]
\tablewidth{0pt}
\tablecaption{Orbital parameters obtained from fitting J0950's radial velocity curves corresponding to the measurements given in Table~\ref{tab:rel_shifts}}
\renewcommand{\arraystretch}{1.3}
\label{tab:model_params}
\tablehead{
\multicolumn{1}{l}{Parameter} & \colhead{Model H$\beta$ $^a$} & \colhead{Data H$\beta$ $^a$} 
}
\startdata
\multicolumn{3}{l}{\underline{\it Free Parameters of the Model}} \\
Period, $P$ (yr)$^b$ & $33^{+7}_{-5}$ & $35^{+6}_{-6}$ \\
Time of conjunction, $T_c$ & $2061^{+9}_{-7}$ & $2063^{+8}_{-7}$ \\
Natural Log of Semi-amplitude, $\ln \left(K/{\rm km\,s^{-1}}\right)$ & $7.88^{+0.47}_{-0.24}$ & $7.80^{+0.37}_{-0.18}$ \\ 
$\sqrt{e}\,\mathrm{cos}\,\omega$ $^c$ & $0.57^{+0.12}_{-0.12}$ & $0.58^{+0.09}_{-0.09}$ \\
$\sqrt{e}\,\mathrm{sin}\,\omega$ $^c$ & $0.55^{+0.10}_{-0.12}$ & $0.49^{+0.11}_{-0.13}$ \\
\noalign{\vskip 6pt}
\multicolumn{3}{l}{\underline{\it Derived Parameters}} \\
Eccentricity, $e$ & $0.65^{+0.13}_{-0.13}$ & $0.58^{+0.12}_{-0.10}$ \\
Tangent of argument of periapsis, $\tan\omega$ & $0.94^{+0.41}_{-0.29}$ & $0.82^{+0.37}_{-0.26}$ \\
Velocity semi-amplitude, $K$ (km s$^{-1}$) & $2640^{+1590}_{-560}$ & $2440^{+1090}_{-400}$ \\
\noalign{\vskip 6pt}
Minimum Reduced $\chi^2$ $^d$ & 0.2 & 0.3 \\
\enddata
\tablecomments{\\
$^a$ All uncertainties correspond to 68\% confidence intervals.\\
$^b$ {\color{black}The period reported here is in the frame of the observer.}\\
$^c$ Finding eccentricity $e$ and argument of periapsis $\omega$ by fitting $\sqrt{e}\,\mathrm{cos}\,\omega$ and $\sqrt{e}\,\mathrm{sin}\,\omega$ is computationally more efficient in \texttt{radvel} than fitting $e$ and $\omega$ directly. \\
$^d$ Reduced $\chi^2$ with 7 degrees of freedom; {\color{black}a small value due to error bars increased by jitter.}} 
\end{deluxetable*}

\subsection{Exploration of the Supermassive Binary Black Hole Scenario and Discussion}\label{subsec:explore_sbhb}
A lower limit on the total mass of the SBHB in J0950 was initially estimated at
$\sim10^6$ M$_\odot$ by \cite{runnoe2017large}.
We obtained limits on the mass assuming that the 
bolometric luminosity does not exceed 
the Eddington luminosity. The bolometric luminosity was estimated by taking the luminosity of J0950's 2002 spectrum––the brightest state of this object––at 5100~\AA, and applying a bolometric correction factor of 10.3 (\citealt{richards2006spectral}). 
This yields a mass of $\gtrsim 9\times10^6 \,\mathrm M_{\odot}$.
We estimated another lower mass limit, taking maximum inclination ($\sin i=1$) and a mass ratio of $q=0$, in the following Keplerian velocity equation,
\begin{equation}
\begin{split}
        M = 3.78\times10^5 & 
    \left[\dfrac{(1+q) \sqrt{1-e^2}}{\sin i}\right]^3 \\ & 
    \left(\dfrac{P}{10\,\mathrm{yr}}\right) 
    \, \left(\dfrac{K}{10^3\, \mathrm{km\,s^{-1}}}\right)^3
    \; \mathrm{M_{\odot}}.
    \end{split}
\label{eq:mass_bh} 
\end{equation}
This equation was derived under the assumption of an unequal mass binary system, in which the less massive, or secondary, BH 
is active and its BLR emits the observed H$\beta$ line (see arguments in \citealt{eracleous2012large,runnoe2015large}).
Using all combinations explored by the MCMC algorithm for the period $P$, the velocity semi-amplitude $K$, and eccentricity $e$, 
the inferred 68\%-confidence lower limit on total mass is $6\times10^6 \,\mathrm M_{\odot}$ and the 99\% lower limit is $3\times10^6 \,\mathrm M_{\odot}$. These inferred masses do not necessarily correspond to the lower limits on $P$, $K$, and $e$. 
All the lower mass limits above are reasonable in the sense that they are less than $10^{10} \,\mathrm M_{\odot}$, a threshold regarded as unphysical. This threshold comes about because the most massive BHs measured are of the order of $10^{10} \,\mathrm M_{\odot}$ and because BHs are not expected to accrete material rapidly enough to surpass this mass (e.g., \citealt{thomas201617,inayoshi2016there,mcconnell2012dynamical}). In line with this argument, the minimum possible inclination for J0950 is $i\approx4^\circ$. 
 Using the distributions of period and mass, applying Kepler's law gives 68\% and 99\% lower limits on the orbital semi-major axis of 0.008~pc and 0.006~pc respectively.

If J0950 is a binary, we attribute the offset broad H$\beta$ emission lines to a BLR associated with the lower mass, or secondary, BH. 
In this scenario, the tidal interaction with the companion BH may limit the size of the BLR. To explore that possibility, we estimate the ``characteristic radius'' of the BLR\footnote{This is the length that corresponds to the observed time delay, $\tau$ between fluctuations of continuum and a broad line, $R_{\rm BLR}\equiv~c\,\tau$.} using the empirical relation from \cite{bentz2013} between the characteristic radius 
and AGN luminosity at 5100~\AA,
\begin{equation}
\log \left(\dfrac{R_{\rm BLR}}{\text{1\;lt-day}}\right)
    =A + B\;
    \log\left[
    \dfrac{\lambda L_{\lambda}\text{(5100\;\AA)}}{10^{44}\;\mathrm{erg\;s^{-1}}} 
    \right]   
    \label{eq:blr_size}
\end{equation}
with $A=1.527\pm 0.031$ and $B=0.533^{+0.035}_{-0.033}$. We applied this relation even though 
it is appropriate only for a single BH and does not account for the dynamical effects in a binary system to get 
a characteristic radius of the BLR 
of $0.02-0.03$~pc.
This value is somewhat higher than
our inferred lower limit on the binary semi-major axis of $\sim10^{-2}$~pc. The extent of the line-emitting region in the BLR is possibly 4$-$5 times greater than $R_{\rm BLR}$ (see discussion in \citealt{runnoe2015large}). Since we only have a lower limit on the orbital semi-major axis, there appears to be no conflict between this separation and the characteristic radius of the BLR.

If J0950 does indeed harbor a binary,  extrapolating from the fit to
its radial velocity curve, it should start turning over soon. 
It is therefore critical that we observe this object in the near future, at a rate of at least one spectrum per year to see if this prediction is borne out. 
It is also important to note that our implementation of radial velocity jitter discussed in Section~\ref{subsec:rv_curve} is simple, and more sophisticated prescriptions of jitter should be developed and employed. 
The description of jitter would impact the uncertainties on the orbital parameters that are inferred from fitting a radial velocity curve. Regardless of whether or not J0950 is a binary, it still displays an exceptional behavior that warrants detailed study.

\subsection{On the Detectability of Doppler Boosting in the Light Curves}\label{subsec:doppler_boost}
{\color{black}The light curves of J0950 exhibit neither substantial nor periodic variability, as discussed in Section \ref{sec:lightcurves}. In addition to the tests in that section, we phase-folded the light curves using the inferred 33~yr period of J0950, binned the magnitude values by phase, and calculated the median magnitudes and spread within each phase bin. We fit the phase-folded, binned light curves with a flat line corresponding to the mean of each light curve, and found reduced $\chi^2$ values $\lesssim$~5, consistent with no strong periodic structure.
Nonetheless, we sought to estimate whether Doppler boosting could produce variability signatures that are either large enough to eventually emerge in the light curves, or too subtle that they remain undetectable. 

Considering a SBHB system, in which a BH carrying the BLR beams light as it orbits, the observed specific flux depends on the angle $\theta$ between the beam's velocity vector and the observer's line of sight, through the relativistic Doppler factor $D=1/\gamma(1-\beta\mu)$, where $\beta=v/c$ and $\mu=\cos\theta$. For a single source, the observed flux scales as $f_{\nu}\propto D^{3-\alpha}$, where $\alpha$ is the spectral index \citep[see, for example,][]{dorazio2023review}. 



Beaming is most pronounced when the binary is in quadrature. In such a configuration the angle $\theta$ is determined only by the inclination of the binary orbit so that $\beta\mu=\pm K/c$, depending on whether the active black hole is moving towards or away from the observer ($K$ is the velocity semi-amplitude obtained from the fit to the radial velocity curve). We can therefore express the ratio of the maximum to minimum observed flux as, 
\begin{equation}
    \dfrac{f_{\nu\mathstrut}^{max}}{f_{\nu\mathstrut}^{min}}=\dfrac{D_{min}^{3-\alpha\mathstrut}}{D_{max}^{3-\alpha\mathstrut}}=\left(\dfrac{1-\beta\mu_{min}}{1-\beta\mu_{max}}\right)^{3-\alpha}=\left(\dfrac{c+K}{c-K}\right)^{3-\alpha}
\end{equation}
Adopting $K=2600$~km~s$^{-1}$ and $\alpha\approx-0.87$ from our spectral fits,  $f_{\nu\mathstrut}^{max}/f_{\nu\mathstrut}^{min}\sim1.07$. 
This corresponds to a magnitude difference of $\sim0.07$––
a very small variation, far below the level of variability seen in the light curves of this object. Thus, in the interpretation of J0950 as a binary, the expected Doppler boosting signatures are likely buried in noise and too weak to be confidently detected. Still, the observed variability does not contradict the binary model.
}

\begin{figure*}[htb!]
  \centering
  \includegraphics[scale=0.55, angle=0]{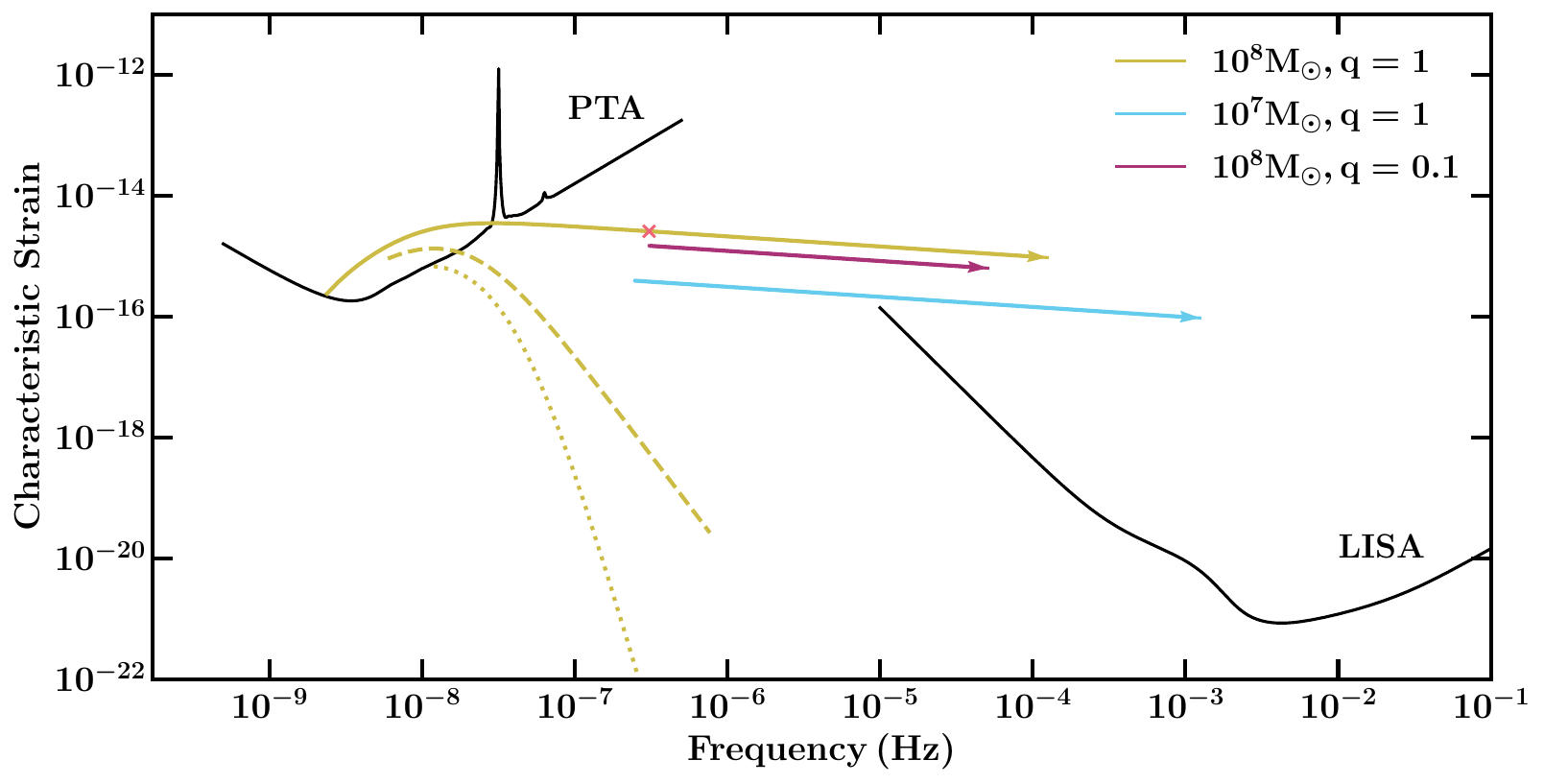}
  \caption{An illustration of the evolution of \bluetext{a system like J0950} across the PTA and LISA bands. For an equal-mass binary of total mass $10^8$M$_{\odot}$, the yellow lines show the evolution track from an eccentricity of 0.65 to approximately zero 
  for harmonics of $n=10$ (dotted), $n=5$ (dashed), and $n=2$ (solid). The red X marks the time by which the eccentricity effectively decays to zero. 
  The solid arrows represent the trajectory for $n=2$ from the point of circularization 
  to merger {\color{black}($\sim$~55~Myr)}. For an unequal-mass binary with a mass ratio $q=0.1$, the track is represented by the purple arrow. The cyan arrow traces the trajectory from circularization to merger for a lower mass binary of $10^7$M$_{\odot}$ with $q=1$. The PTA and LISA sensitivity curves were plotted using the \texttt{hasasia} package (specifically its module for deterministic sensitivity curves without the GW background, providing a best-case, theoretical estimate; \citealt{hazboun2019}) and the LISA sensitivity calculator by \cite{robson2019}.
}
  \label{fig:gw_curves}
\end{figure*}

\subsection{Speculation: Evolution of J0950 in the GW landscape}\label{subsec:gw_speculation}
Given our findings above, it is interesting to speculate about the expected evolution of the GW signal from \bluetext{a system like} J0950 in the characteristic strain versus frequency diagram.
Figure~\ref{fig:gw_curves} illustrates the path of \bluetext{such a system} as its eccentricity evolves from 0.65 to approximately zero, assuming that GW emission is the dominant mechanism driving orbital evolution. 
This assumption does not account for the ongoing accretion in \bluetext{the system}, which implies continued interaction with the surrounding gas. Gas dynamics could influence the timescale of orbital decay and may sustain a non-zero eccentricity for an extended period. 
Figure~\ref{fig:gw_curves} shows three different harmonics of the GW signal.
The steps involved in producing the components of the figure are outlined in Appendix~\ref{appendixB}. {\color{black} Here, we predict the overall evolution of \bluetext{the system} across the PTA and LISA bands using theoretical sensitivity curves by \cite{hazboun2019} and \cite{robson2019}, respectively}.
For the purposes of this illustration we adopt representative values of the mass and mass ratio, as indicated in Figure~\ref{fig:gw_curves}.  

Initially, the characteristic strain, $h_c$, 
of each higher-order harmonic, $n=5$ and $n=10$, is strong. However, as the orbit circularizes, the $n=2$ harmonic becomes dominant, while the other harmonics decay quickly, albeit at different rates. As the eccentricity approaches zero, the $n=2$ becomes the only significant harmonic. To evolve \bluetext{the system} further, up to the merger, we take the point of coalescence as approximately the separation at the innermost stable circular orbit (ISCO; the last stable circular orbit beyond which the BHs will merge). For a non-spinning (Schwarzschild) BH, the ISCO separation 
is given by $a=6GM_1/c^2$ (where $M_1$ is the mas of the more massive, or primary, BH). The path of \bluetext{the system} from circularization of the orbit to ISCO, for varying configurations of total mass and mass ratios, is indicated by arrows in Figure~\ref{fig:gw_curves}, with the tip of the arrows representing the merger. 



Finally, we estimate the evolutionary time scales for the GW signal from \bluetext{the system}, adopting a characteristic binary mass of $\sim$10$^8\,\rm M_{\odot}$. Using the binary decay lifetime from \cite{peters1964}, 
it will take 55~Myr for this object to merge. The details of this calculation are given in Appendix~\ref{appendixB}. 
The time \bluetext{the system} will spend in the PTA frequency band spans from when the $n=5$ harmonic (initially the strongest at \bluetext{the system's} current eccentricity, $e=0.65$) first enters the band, 
to the point where the $n=2$ harmonic intersects the PTA sensitivity curve. \bluetext{The system} will spend $\sim$ 54~Myr in this band. Furthermore, it will take a time comparable to its overall decay lifetime of 55~Myr for it to evolve from its current eccentricity to the low-frequency end of the LISA band. Once in the LISA band, it will merge within $\sim$1 month.

It is clear from the exercise above that, if J0950 is a binary, it might 
eventually be detectable by PTA experiments. When it merges at the end of its evolution, 
it may be detectable by experiments similar to LISA.
Therefore, J0950 could serve as an example of the 
systems producing the stochastic background detected by PTAs. Moreover, if there are other low-redshift systems \bluetext{with properties similar to those adopted above} that are now emitting in the PTA band, we will have a chance of detecting them in the near future.




\section{Conclusions}\label{sec:conclusions2}
{\color{black}The spectroscopic supermassive black hole binary candidate J0950$+$5128 (abbreviated J0950) was identified in 2012 by the large velocity shift of its broad H$\beta$ line. Continued spectroscopic monitoring spanning 22 years, using the SDSS, Palomar Hale, Keck, and Hobby-Eberly telescopes, has revealed a systematic $\sim$3500~km~s$^{-1}$ shift in the broad H$\beta$ emission line over time, including a reversal in the sign of the velocity. Using a cross-correlation technique, we measured the velocity shifts in both the observed spectra and their model reconstructions, including versions with the isolated broad H$\beta$ component after spectral decomposition. All measurement approaches indicate a monotonic increase from negative to positive velocities.

Through tests and simulations, we identified sources of uncertainty of the same magnitude as those from cross correlation, including those arising from the subtraction Fe~\textsc{ii} during spectral decomposition, as well as from the increasing width of the broad H$\beta$ line between observations. The largest and dominant source of uncertainty is jitter, which we combined with our measurement uncertainties to account for the intrinsic stochastic variability of quasars. Alongside the spectroscopic analyses, we examined the available light curves from CRTS, NEOWISE, and ZTF, and found no evidence of periodicity. 

We considered several interpretations for the observed changes in broad emission line of J0950 and disfavored BLR obscuration by a dust cloud, outflows, and a recoiling black hole. The possibility that a SBHB or a perturbed, disk-like BLR could produce the broad emission line shifts remain viable. The latter interpretation requires additional data and specialized modeling for further tests that we defer to future work.

To evaluate the SBHB interpretation further, we fit the observed radial velocity curve with an orbital model. The fit yields a period of 33~years (in the frame of the observer) and an eccentricity of 0.65. We estimate lower limits on the semi-major axis and black hole mass of approximately 10$^{-2}$~pc and $10^7\;{\rm M}_\odot$, respectively. Under the binary interpretation of J0950, we used the orbital parameter estimates to explore whether Doppler boosting signatures could be detected in the light curves, finding that such signatures would be too weak to observe. We also used the orbital parameters to explore the evolution of a hypothetical GW signal from J0950. If the system is indeed a binary, its current expected signal falls in the detection band of PTA experiments, while its eventual merger signal would fall within the LISA detection band. We conclude that the quasar J0950 is an exceptional object. Regardless of whether it is a supermassive binary candidate, it warrants extensive monitoring to better characterize its properties and determine its true nature.
}

\section*{Acknowledgements}
{\color{black}We thank the anonymous referee for their detailed comments that helped us improve the analysis of the data and presentation of the results.} N.N.M acknowledges the support of the Penn State Science
Achievement Graduate Fellowship, and helpful discussions with Laura Duffy, Meghan Delamer, Lucas Brefka, and Evan Fitzmaurice at Penn State. M.E. and N.N.M acknowledge support from the National Science Foundation under grant AST-2205720. J.C.R. and C.D. acknowledge support from the National National Foundation (NSF) under grant AST-2205719. T.B. acknowledges support from the National Science Foundation under grant AST-2307278 and from the Research Corporation for Science Advancement under award CS-SEED-2023-008. J.S is supported by an NSF Astronomy and Astrophysics Postdoctoral Fellowship under award AST-2202388, and acknowledges previous support from the JPL RTD program. M.C. is funded by the European Union (ERC-StG-2023, MMMonsters, 101117624). The work of D.S. and J.L. was carried out at the Jet Propulsion Laboratory, California Institute of Technology, under a contract with NASA. We thank S. George Djorgovski for allowing us to use some of his time allocation on Keck to obtain spectra of J0950 on 04/13/2021.

The Low Resolution Spectrograph 2 (LRS2) was developed and funded by the University of Texas at Austin McDonald Observatory and Department of Astronomy, and by Pennsylvania State University. We thank the Leibniz-Institut fur Astrophysik Potsdam (AIP) and the Institut fur Astrophysik Goettingen (IAG) for their contributions to the construction of the integral field units. We acknowledge the Texas Advanced Computing Center (TACC) at The University of Texas at Austin for providing high performance computing, visualization, and storage resources that have contributed to the results reported within this paper.

Funding for the SDSS has been provided by the Alfred P. Sloan Foundation, the Participating Institutions, the National Science Foundation, the U.S. Department of Energy, the National Aeronautics and Space Administration, the Japanese Monbukagakusho, the Max Planck Society, and the Higher Education Funding Council for England. The SDSS Web Site is http://www.sdss.org/.

The SDSS is managed by the Astrophysical Research Consortium for the Participating Institutions. The Participating Institutions are the American Museum of Natural History, Astrophysical Institute Potsdam, University of Basel, University of Cambridge, Case Western Reserve University, University of Chicago, Drexel University, Fermilab, the Institute for Advanced Study, the Japan Participation Group, Johns Hopkins University, the Joint Institute for Nuclear Astrophysics, the Kavli Institute for Particle Astrophysics and Cosmology, the Korean Scientist Group, the Chinese Academy of Sciences (LAMOST), Los Alamos National Laboratory, the Max-Planck-Institute for Astronomy (MPIA), the Max-Planck-Institute for Astrophysics (MPA), New Mexico State University, Ohio State University, University of Pittsburgh, University of Portsmouth, Princeton University, the United States Naval Observatory, and the University of Washington.

Funding for SDSS IV has been provided by the Alfred P. Sloan Foundation, the U.S. Department of Energy Office of Science, and the Participating Institutions. SDSS-IV acknowledges support and resources from the Center for High Performance Computing  at the University of Utah. The SDSS website is www.sdss4.org.

SDSS-IV is managed by the Astrophysical Research Consortium for the Participating Institutions of the SDSS Collaboration including the Brazilian Participation Group, the Carnegie Institution for Science, Carnegie Mellon University, Center for Astrophysics | Harvard \& Smithsonian, the Chilean Participation Group, the French Participation Group, Instituto de Astrof\'isica de Canarias, The Johns Hopkins University, Kavli Institute for the Physics and Mathematics of the Universe (IPMU) / University of Tokyo, the Korean Participation Group, Lawrence Berkeley National Laboratory, Leibniz Institut f\"ur Astrophysik Potsdam (AIP),  Max-Planck-Institut f\"ur Astronomie (MPIA Heidelberg), Max-Planck-Institut f\"ur Astrophysik (MPA Garching), Max-Planck-Institut f\"ur Extraterrestrische Physik (MPE), National Astronomical Observatories of China, New Mexico State University, New York University, University of Notre Dame, Observat\'ario Nacional / MCTI, The Ohio State University, Pennsylvania State 
University, Shanghai Astronomical Observatory, United 
Kingdom Participation Group, Universidad Nacional Aut\'onoma 
de M\'exico, University of Arizona, University of Colorado Boulder, 
University of Oxford, University of Portsmouth, University of Utah, University of Virginia, University of Washington, University of Wisconsin, Vanderbilt University, and Yale University.

The NANOGrav collaboration receives support from National Science Foundation (NSF) Physics Frontiers Center award numbers 1430284 and 2020265, the Gordon and Betty Moore Foundation, NSF AccelNet award number 2114721, an NSERC Discovery Grant, and CIFAR. Part of this research was carried out at the Jet Propulsion Laboratory, California Institute of Technology, under a contract with the National Aeronautics and Space Administration.

{\color{black}
The CSS survey is funded by the National Aeronautics and Space Administration under Grant No. NNG05GF22G issued through the Science
Mission Directorate Near-Earth Objects Observations Program. The CRTS survey is supported by the U.S.~National Science Foundation under grants AST-0909182.

This publication makes use of data products from the Near-Earth Object Wide-field Infrared Survey Explorer (NEOWISE), which is a joint project of the Jet Propulsion Laboratory/California Institute of Technology and the University of California, Los Angeles. NEOWISE is funded by the National Aeronautics and Space Administration.

Based on observations obtained with the Samuel Oschin 48-inch Telescope at the Palomar Observatory as part of the Zwicky
Transient Facility project. ZTF is supported by the National Science Foundation under Grant No. AST-1440341 and a collaboration including Caltech, IPAC, the Weizmann Institute for Science, the Oskar Klein Center at Stockholm University, the University of Maryland, the University of Washington, Deutsches Elektronen-Synchrotron and Humboldt University, Los Alamos
National Laboratories, the TANGO Consortium of Taiwan, the University of Wisconsin at Milwaukee, and Lawrence Berkeley National Laboratories. Operations are conducted by COO, IPAC, and UW.

This research has made use of the NASA/IPAC Infrared Science Archive, which is funded by the National Aeronautics and Space Administration and operated by the California Institute of Technology.
}

%

\vspace{5mm}
\facilities{Sloan(SDSS;BOSS), HET(LRS;LRS2), Palomar(DBSP), Keck(LRIS), {\color{black}IRSA, CRTS, NEOWISE, ZTF}}


\software{\texttt{numpy} \citep{numpy_harris2020array}, \texttt{astropy} \citep{astropy2013A&A...558A..33A, astropy2018AJ....156..123A, astropy2022ApJ...935..167A}, \texttt{scipy} \citep{scipy2020SciPy-NMeth}, \texttt{matplotlib} \citep{matplotlib_Hunter:2007}, \texttt{extinction} \citep{barbary_k_2016}, \texttt{radvel} \citep{fulton2018radvel}, \texttt{spectres} \citep{spectres2017arXiv170505165C}, \texttt{pandas} \citep{pandas2022zndo...3509134T} 
          }



\clearpage
\appendix

\section{Side-by-Side Comparison of Models With and Without Noise for the Full Spectra and Isolated Broad H$\beta$ Lines of J0950}\label{appendixA}

\begin{figure} [h]
    \centering
    \includegraphics[width=3.2in]{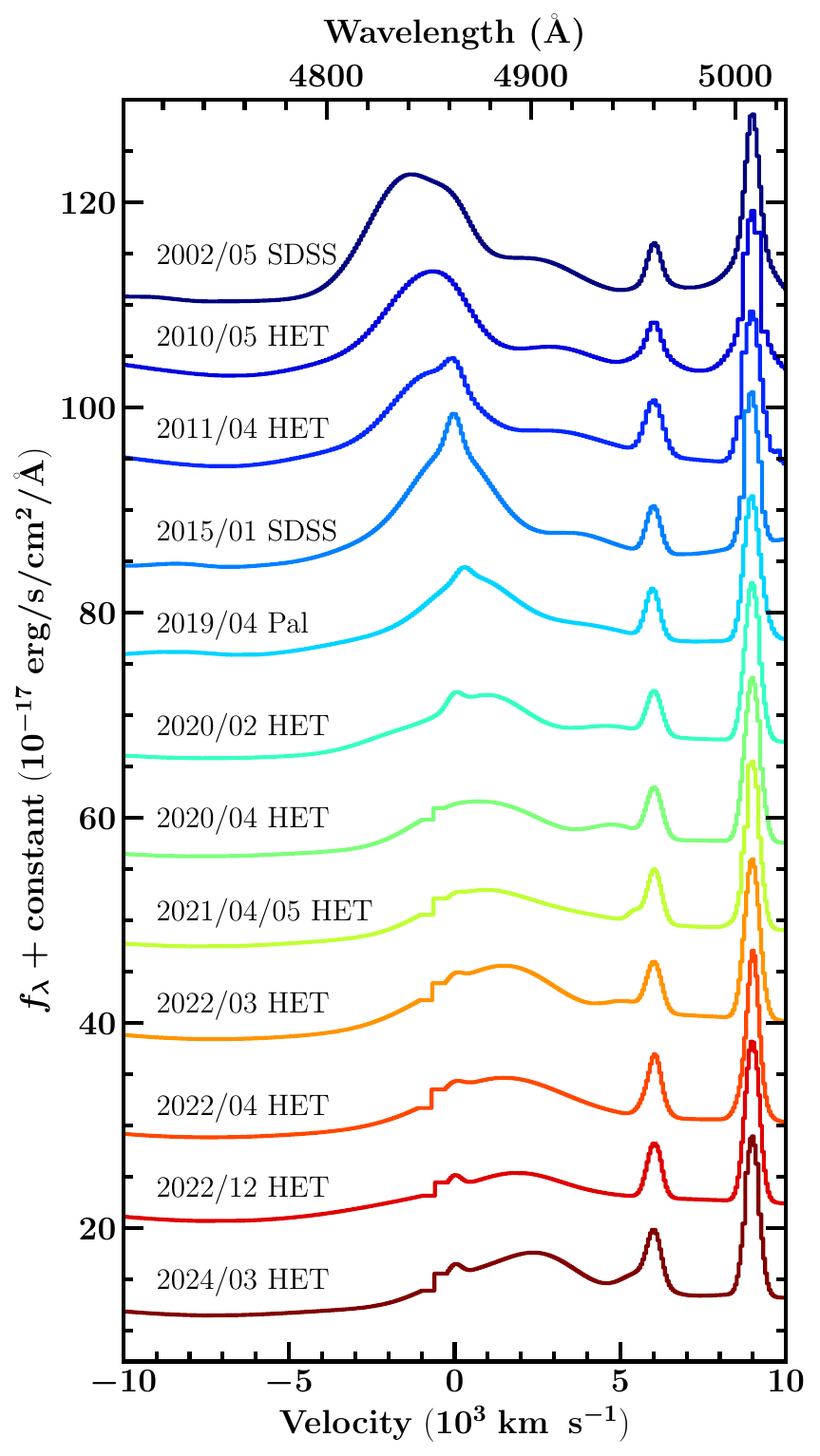} 
    \includegraphics[width=3.2in]{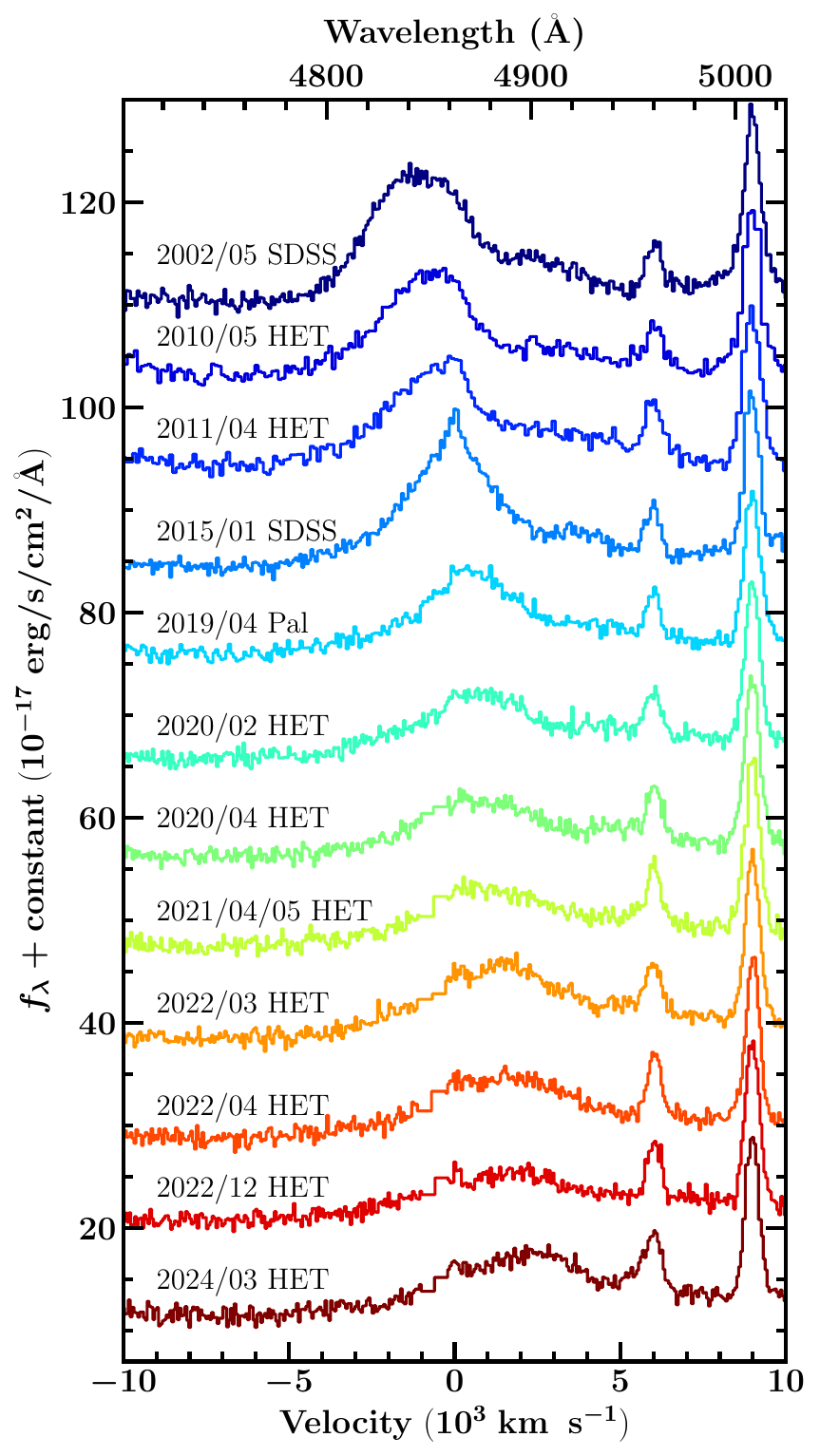}
\caption{
\textbf{\textit{Left}}: Models of the full spectra of J0950 {\color{black} after subtracting the host galaxy contribution, as noted in Section \ref{subsec:crosscor_data}}. {\color{black}The models scaled to match the [O~\textsc{iii}] flux and continuum of the 2002 spectrum and vertically offset for clarity.} \textbf{\textit{Right}}: The models as on the \textit{Left}, but with synthetic noise added, as described in Section~\ref{subsec:crosscor_models}. {\color{black} Refer to the the text of this appendix 
for an explanation of the features in these spectra. See also  Figure~\ref{fig:decomp example} for an annotated illustration of these features.}
} 
\label{fig:full_model}
\end{figure}

\begin{figure} 
    \centering
    \includegraphics[width=3.2in]{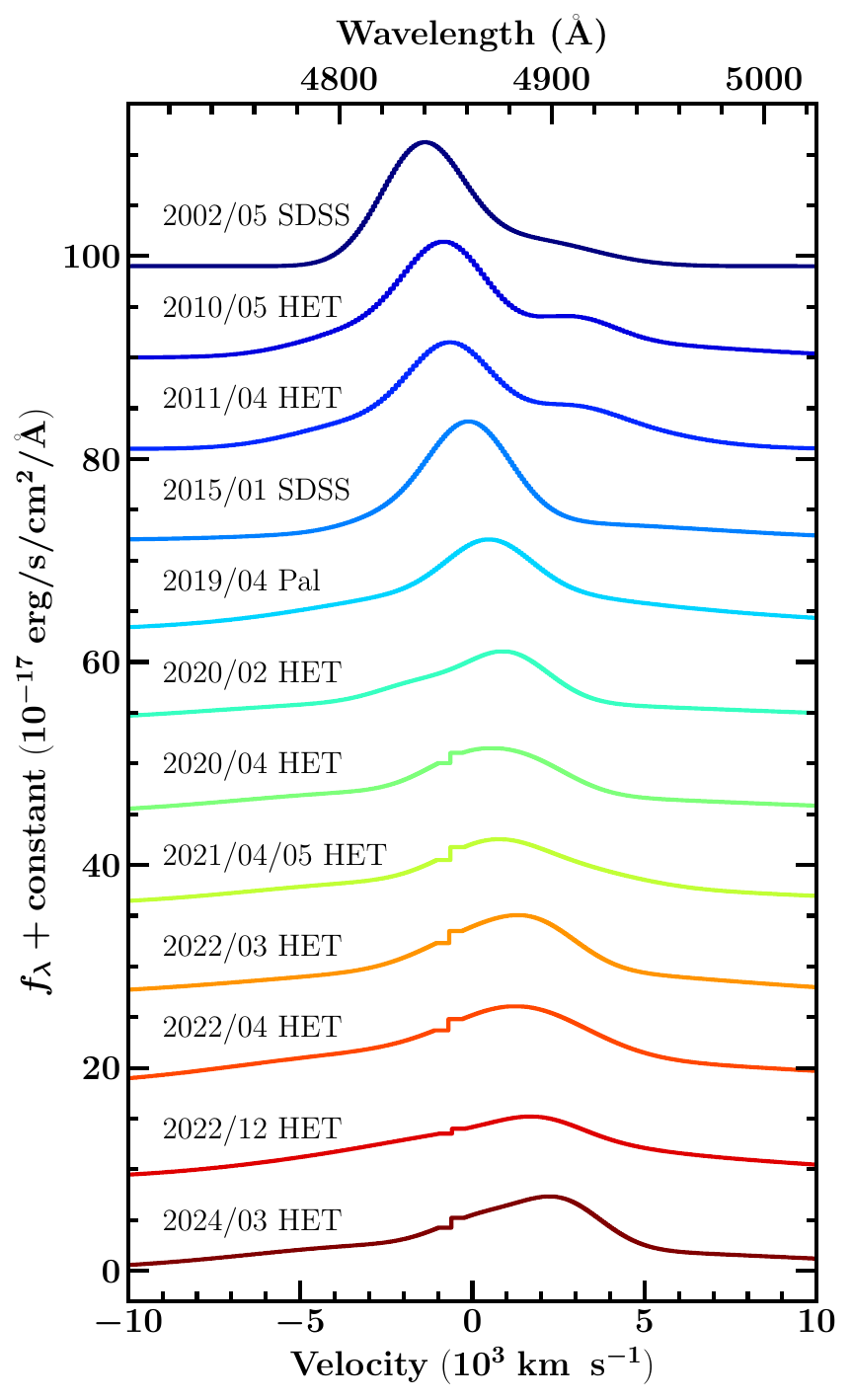}
    \includegraphics[width=3.2in]{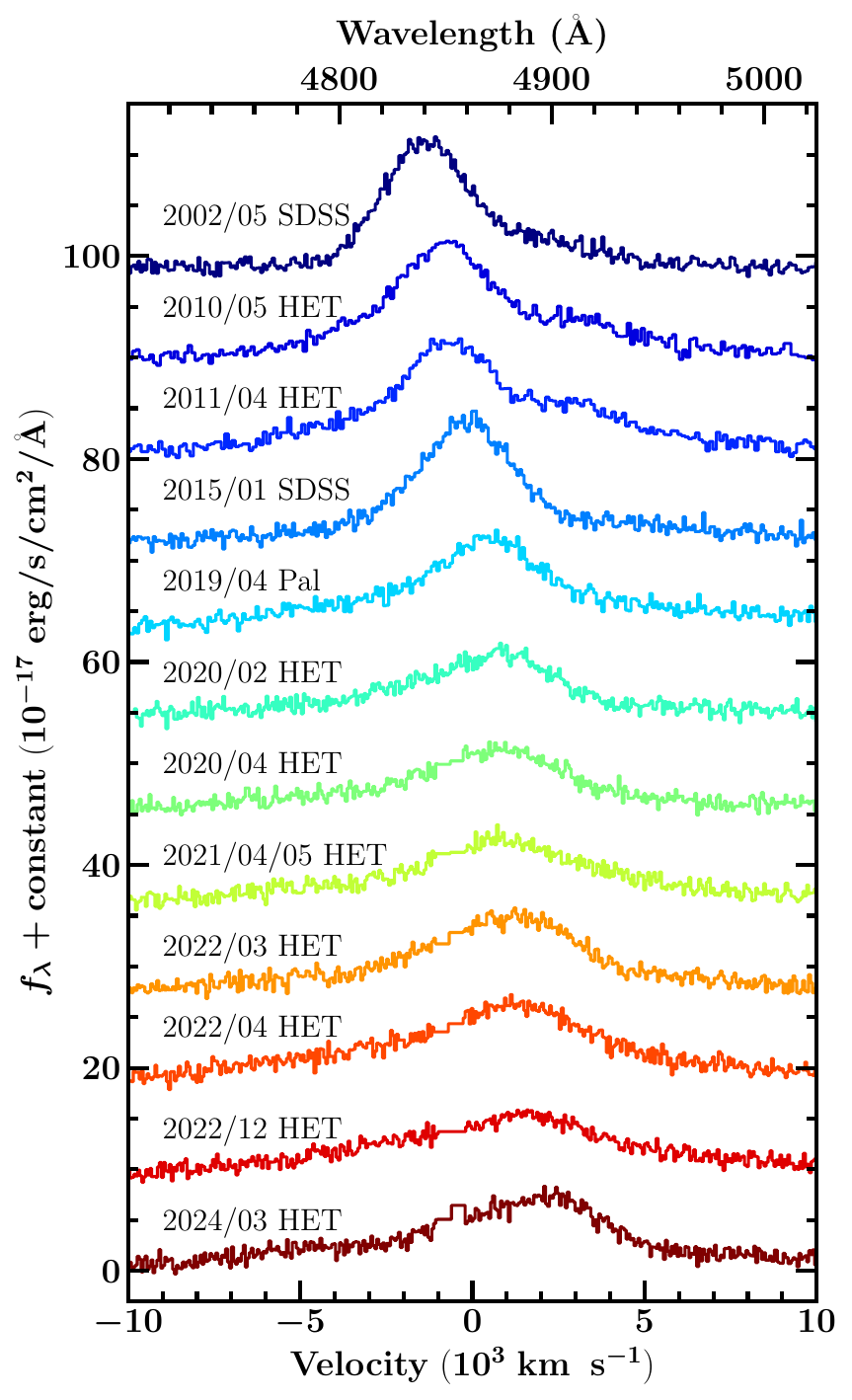}
\caption{\textbf{\textit{Left}}: Models of the isolated broad H$\beta$ emission line of J0950. {\color{black}The models are vertically offset by a constant amount for clarity.} \textbf{\textit{Right}}: The same models as on the \textit{Left}, but with synthetic noise added, as described in Section~\ref{subsec:crosscor_models}. 
{\color{black} 
See also Figure~\ref{fig:decomp example} for an annotated illustration of the broad H$\beta$ line.} 
}
\label{fig:hbeta_model}
\end{figure}

In this appendix, we show spectral decomposition models for the spectra of J0950. Each model in Figure~\ref{fig:full_model} includes the 
quasar continuum, Fe~\textsc{ii} lines, broad and narrow He~\textsc{ii} $\lambda$4687 {\color{black}(a weak line in every spectrum of J0950)}
and H$\beta$ $\lambda$4863
emission lines, and the [O \textsc{iii}] $\lambda\lambda$4959,5007 doublet {(\color{black}see Figure~\ref{fig:decomp example} for an annotated illustration of these features)}, while Figure~\ref{fig:hbeta_model} shows models of the isolated broad H$\beta$ lines. Also presented in both figures are noisy versions of the models. These model spectra are referenced in Section~\ref{subsec:crosscor}, where we quantify sources of random and systematic errors by cross correlating the broad H$\beta$ emission lines in the full and isolated spectral models.




\clearpage
\section{MCMC posterior distributions of orbital parameters from radial velocity fits of J0950}\label{appendix_corner}

\begin{figure}[htbp] 
  \centering
  \includegraphics[width=6.5in, angle=0] 
  {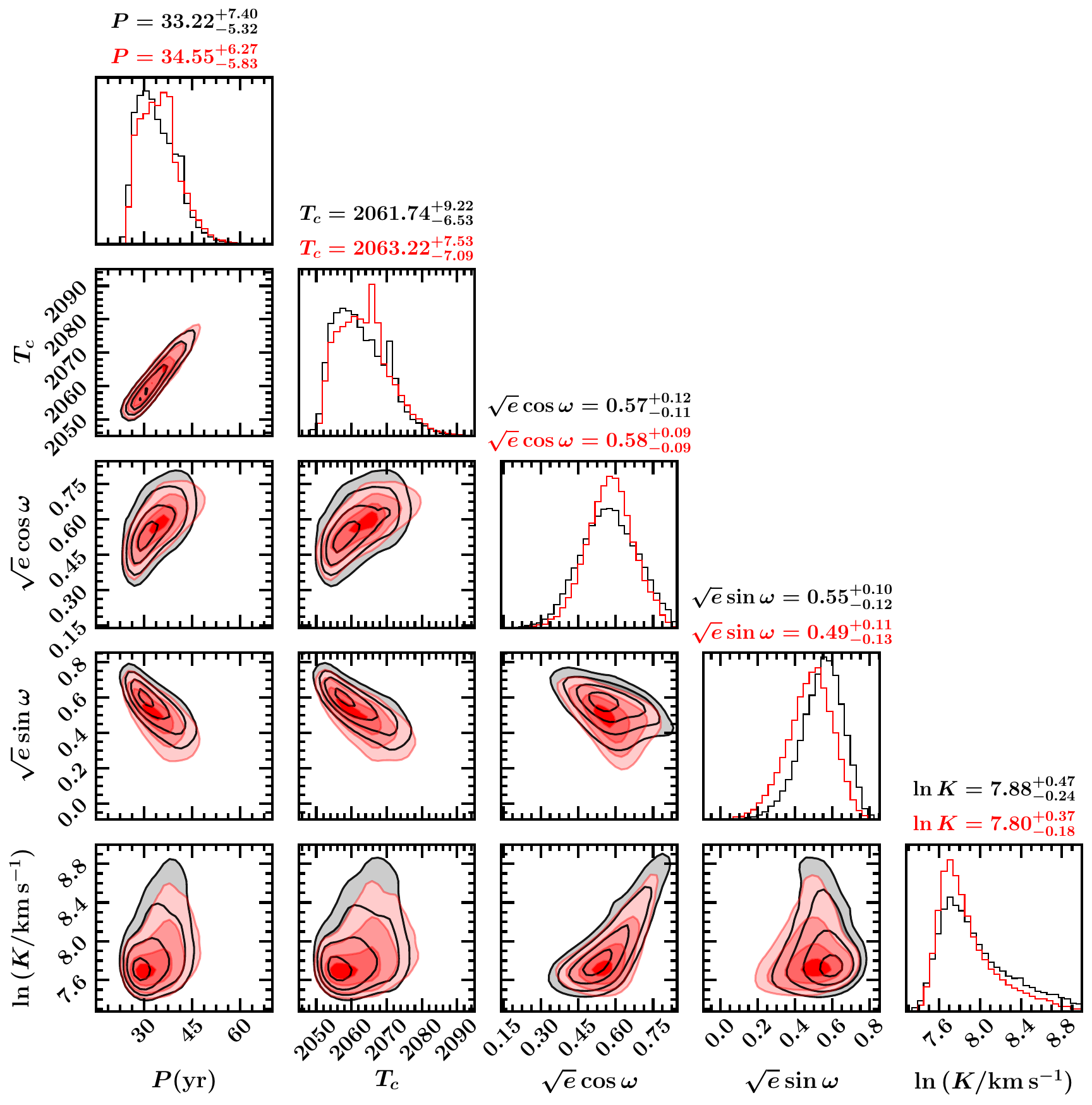}
  \caption{\color{black}Posterior distributions of the orbital parameters for a SBHB model fitted to the radial velocity curves of J0950. The black contours are based on shift measurements from its isolated broad H$\beta$ models, while the red contours are based on shift measurements from the isolated broad H$\beta$ lines of the data. For each parameter, the median values and 68\% confidence limits are shown above the plot.}
  \label{fig:corner_combined}
\end{figure}

{\color{black} In Section \ref{subsec:rv_fitting}, we performed MCMC fits to two radial velocity curves of J0950: one based on the absolute shifts of the isolated broad H$\beta$ models, and another based on the absolute shifts measured from the isolated broad H$\beta$ lines of the data. The fitted orbital parameters are the period ($P$), the time of conjunction ($T_c$), the natural log of the velocity semi-amplitude (ln $K$), and $\sqrt{e}\,\mathrm{cos}\,\omega$ and $\sqrt{e}\,\mathrm{sin}\,\omega$. Figure \ref{fig:corner_combined} shows the corner plot of the posterior distributions from fitting the model-based velocity curve, overlaid with those from fitting the data-based curve.
}

\clearpage
\section{Evolution of \bluetext{a system like J0950} in the PTA and LISA gravitational-wave landscape}\label{appendixB}

The characteristic strain in the detector frame of an eccentric binary at the $n$th harmonic, can be determined using a simplified expression from \citealt{breivik2020},
\begin{equation}
\begin{aligned}
h_{c,n,d}^2 &= \dfrac{2}{3\pi^{4/3}}\dfrac{(G\mathcal{M}_c)^{5/3}}{c^3D_L^2}\dfrac{1}{f^{1/3}_{n,d}(1+z)^2}\left(\dfrac{2}{n}\right)^{2/3}\dfrac{g(n,e)}{F(e)}\\
\end{aligned}
\end{equation}
Here, $\mathcal{M}_c$ is the chirp mass, $D_L$ is luminosity distance, $f_{n,d}$ is the detector frame GW frequency, $F(e)$ is an eccentricity enhancement factor, and $g(n,e)$ is a function involving a series of Bessel functions of the first kind, that captures the contribution of each harmonic $n$ and eccentricity $e$ to the gravitational radiation power spectrum (see \citealt{peters&mathews1963}).  \\
To produce the evolutionary tracks of \bluetext{a system like J0950} for the different harmonics shown in Figure~\ref{fig:gw_curves}, we considered a range of eccentricities from 0.65 to approximately zero. For a given $e$, the corresponding orbital separation $a$ can be obtained using an expression from \citealt{peters1964},
\begin{equation}
\begin{aligned}
a=\dfrac{c_0e^{12/19}}{(1-e^2)}\left[1+\dfrac{121}{304}e^2\right]^{870/2299}
\label{eq:separation}
\end{aligned}
\end{equation}
where $c_0$ is calculated from the instantaneous $a$ and $e$, which \bluetext{we take to be} $a=0.02$~pc (semi-major axis for a binary mass of $\sim$10$^8\,\rm M_{\odot}$ and orbital period of 33 yr) and $e=0.65$. Then for each separation, the GW frequency is,
\begin{equation}
\begin{aligned}
f_{n,d}=n\, \dfrac{1}{2\pi}\left(\dfrac{G\,M}{a^3}\right)^{1/2}
\label{eq:frequency_gw}
\end{aligned}
\end{equation}
The equations above were applied iteratively over a range of eccentricities at a specific $n$. Once the orbit becomes circular, the $n=2$ harmonic dominates. \bluetext{The system} was then evolved from the frequency where the eccentricity is approximately zero to the frequency at merger, maintaining a fixed $n=2$. Recall that for this exercise, we defined ``merger'' as the separation at ISCO in Section~\ref{subsec:gw_speculation}.

The timescales provided in Section~\ref{subsec:gw_speculation} for \bluetext{such a system} to merge and its GW signal to be detected by PTA experiments and LISA, were computed using the decay lifetime equation given by \citealt{peters1964}, 
\begin{equation}
\begin{aligned}
T &= \dfrac{12}{19}\dfrac{c_0^4}{\beta}\int_{e_f}^{e_i}\dfrac{de\,e^{29/19}[1+(121/304)e^2]^{1181/2299}}{(1-e^2)^{3/2}}
\label{decay_lifetime}\\
\beta &= \dfrac{64}{5}\dfrac{G^3M_1 M_2(M_1+M_2)}{c^5}
\end{aligned}
\end{equation}
where $e_i>e_f$.
For example, we found that the time it will take for \bluetext{such a system} to merge, assuming it is an equal-mass $\sim$10$^8\,\rm M_{\odot}$ binary, is 55~Myr. This was determined by starting with an initial eccentricity of $e_i=0.65$ and a final eccentricity $e_f$ corresponding to the separation at ISCO. In Figure~\ref{fig:gw_curves}, the ISCO is indicated by the tip of the arrow of the $n=2$ harmonic. To determine the value of $e_f$, we first took the GW frequency at this tip and calculated the corresponding separation by rearranging Eq.~\ref{eq:frequency_gw}. In this particular example, the ISCO separation can also be directly calculated using $a=6GM_1/c^2$. Then, we obtained $e_f$ by numerically evaluating $e$ in Eq.~\ref{eq:separation}. Finally, the time to evolve from $e_i$ to $e_f$ was calculated via numerical integration. 

\clearpage



\clearpage
\bibliography{bibfile}{}
\bibliographystyle{aasjournalv7}



\end{document}